\documentclass[preprint,aps,amsmath,amssymb,prb,superscriptaddress,longbibliography%
]{revtex4-2}
\usepackage{graphicx}
\usepackage{bm}
\usepackage{amsmath}
\usepackage[utf8]{inputenc}
\usepackage[T1]{fontenc}
\usepackage{mathptmx}
\usepackage{physics}
\usepackage{float}
\usepackage{hyperref}
\usepackage{xcolor}
\usepackage{soul}
\usepackage{comment}
\usepackage{array}
\makeatletter
\newcounter{subsubsubsection}[subsubsection]

\newcommand\subsubsubsection{\@startsection{subsubsubsection}{4}{\z@}%
  {-3.25ex\@plus -1ex \@minus -.2ex}%
  {1.5ex \@plus .2ex}%
  {\normalfont\normalsize\itshape}}
\newcommand\l@subsubsubsection{\@dottedtocline{4}{7em}{4em}}

\newcommand{\hH}{{\hat{H}}}
\newcommand{\hR}{{\hat{R}}}
\newcommand{\hP}{{\hat{P}}}

\newcommand{\hY}{{\hat{Y}}}
\newcommand{\hLambda}{{\hat{\Lambda}}}
\newcommand{\hL}{{\hat{L}}}
\newcommand{\hU}{{\hat{U}}}
\newcommand{\hD}{{\hat{D}}}

\newcommand{\hGamma}{{\hat{\Gamma}}}

\newcommand{\hbGamma}{{\hat{\bm \Gamma}}}

\newcommand{\dpb}[2]{ \left\{\left\{ #1, #2 \right\}\right\}}

\newcommand{\acom}[2]{\left[ #1, #2 \right]_+}
\newcommand{\com}[2]{\left[ #1, #2 \right]_-}

\date{July 2026}
\begin{document}

\title{An Exact Formulation of Phase-Space Electronic Structure Theory in One Dimension by Exploiting the Zero Curvature Condition}
\author{Zain Zaidi}
\email{zz4271@princeton.edu}
\affiliation{Department of Chemistry, Princeton University, Princeton, New Jersey 08540, United States}
\author{Yotam M.\ Y.\ Feldman}
\email{yotam.feldman@gmail.com}
\affiliation{Department of Chemistry, Princeton University, Princeton, New Jersey 08540, United States}
\author{Linqing Peng}
\email{lp9673@princeton.edu}
\affiliation{Department of Chemistry, Princeton University, Princeton, New Jersey 08540, United States}
\author{Joseph E. Subotnik}
\email{subotnik@princeton.edu}
\affiliation{Department of Chemistry, Princeton University, Princeton, New Jersey 08540, United States}

\begin{abstract}
  We show that, for the unique case of a system of electrons and nuclei in one dimension, the $\hGamma$ operator in phase space electronic structure theory has a zero non-Abelian curl. Using this fact, we are able to show that a phase space electronic structure framework in Wigner space has a one-to-one mapping to the exact fully quantum vibronic Hamiltonian. In other words, one loses no information by working within the framework of phase space electronic structure theory rather than Born-Huang theory; the only difference is that, whereas Born-Huang theory represents the total vibronic Hamiltonian as a quadratic order expansion in $\hbar$, the Hamiltonian becomes an infinite order expansion in $\hbar$ within a phase space electronic structure framework (a so-called "Generalized Born-Huang" approach). That being said, numerical results at first or second order demonstrate that, for a limited number of electronic states, phase space electronic structure theory strongly outperforms  Born-Huang theory -- a result that justifies a great deal more research into this novel electronic structure approach.
\end{abstract}
\maketitle

\section{Introduction: Nonadiabatic Approaches to Chemistry Through Wigner Transformations}

The fundamental difference between nuclei and electrons is their mass: because nuclei are far heavier than electrons, chemists and physicists almost always treat nuclei and electrons at different levels of theory.  The usual means by which this difference manifests is the famous Born-Huang representation\cite{born-huang_dynamical_1955}.  According to BH theory, one generates a basis of electronic states $\left\{\ket{\Phi_i (R)} \right\}$ and energies $E_i(
R)$  by freezing the nuclei (at position $R$) and then solving the electronic structure problem:
\begin{align}
    \hH_{el}(R) &= \frac{\hat{p}^2}{2m} + \hat{V}(\hat{r};R) \\
    \hH_{el}(R) \ket{\Phi_i (R)} &= E_i(R)\ket{\Phi_i  (R)} 
\end{align}
The full, nuclear+electronic Schrodinger equation is then projected into this basis, and the result is usually written down as: 

\begin{align}\label{eq:BH}
    \hH = \sum_{i,j} \ket{i} \left( \sum_k \frac{\left( \hat{P} \delta_{ik} - i\hbar \hat{d}_{ik} \right)\left( \hat{P}\delta_{kj} - i\hbar \hat{d}_{kj} \right)}{2M} + E_{i}(\hat{R}) \delta_{ij}\right) \bra{j}
\end{align}
Here, the matrix elements that connect different BH states are the derivative couplings\cite{yarkony:1996:rmp},
\begin{align}
    \hat{d}_{ij} = \bra{i} \frac{\partial}{\partial R} \ket{j}
\end{align}
(As a matter of convenience, we will use  the notation $\ket{\Phi_i (R)}$ or  $\ket{i}$ equivalently to refer to the same $i^{th}$ electronic state.) 
As long as one employs a complete set of electronic states, the BH expansion above is exact. In practice, however, one can rarely (if ever) include a complete, infinite set of electronic states. Thus, instead one usually employs a handful of states. In fact, for the majority of chemical physics problems, one focuses on only the ground state. Indeed, the so-called Born-Oppenheimer (BO) approximation\cite{cederbaum:review:conicalbook} stipulates that all nuclear dynamics and vibrations proceed along the ground-state potential energy surface $V_0(R)$, which is the basis for all current liquid and biological molecular modeling\cite{allentildesleybook} and most structural simulations of materials\cite{martinbook} as well.

Now, it is crucial to emphasize that the BH framework is not the only possible framework for chemical dynamics. In fact, it should be obvious to the reader that,  so long as one chooses a complete set of electronic states, one can solve the total, coupled nuclear-electronic Schrodinger equation in any basis. For instance, within conventional chemistry, besides the adiabatic basis, one often discusses the notion of ``diabatic bases,'' \cite{cave:1996:gmh,cave:1997:gmh,subotnik:2008:boysgmh,voityuk:2002:fcd,hsu:2006:tt,subotnik:2015:acr} i.e. basis sets $\left\{ \ket{\Phi_i} \right\}$ which should have constant character and are weak functions of  nuclear position,  $\left\{ \ket{\Phi_i(R)} \right\}$ . In fact,  any position-dependent electronic basis can be plugged into the total Schrodinger equation and the result is always of the form in  Eq. \ref{eq:BH} always holds for any set of electronic states -- though $E_i(R)$ need not be diagonal; in other words, even though the BH expansion is formally designed around diagonalizing $\hH_{el}$, the equations of motion are quite general for any  position-dependent electronic basis.

More generally, however, if one were to go beyond BH theory (and one wants to preserve the notion of potential energy surfaces without invoking an exact factorization\cite{gross:2015:jcp:vcd_exact_factorization}, one can imagine parameterizing the electronic states by both nuclear position $R$ {\em and} nuclear momentum $P$, which is the essence of phase space electronic structure theory\cite{xuezhi:cpr:review:2026} (PSEST). Here one  diagonalizes $H_{PS}(R,P)$ instead of $H_{el}(R)$, where
\begin{align}\label{eq:PS-intro-hamiltonian}
    \hH_{PS}(R,P) = \frac{(P-i\hbar \hGamma)^2}{2M}
 + \hH_{el}(R)
 \end{align}
Here, we include an electronic operator $\hGamma$ that couples nuclear and electronic momentum (see Sec. \ref{Sec:PS-LF}). In order to construct the PS electronic Hamiltonian, however, one must first specify what one means by $R$ and $P$. After all, formally, these would appear to be nuclear operators that cannot be specified at the same time according to the Heisenberg uncertainty relationship.  Thus, in order to understand and derive PS electronic structure theory, one must formally invoke partial Wigner transformations\cite{kapral:1999:jcp,kapral:2010:jcp_pbme,ymrhee:2014:jcp_pbme_new}, which transform nuclear operators to symbols.
To that end, for use below, let us formally define the Wigner transform $W$ of an operator $\hat{O}$ as\cite{tannor:quantumbook}:
\begin{equation}
    \hat{O}_W(R,P) = \int dR' \left< R + \frac{R'}{2} \right| \hat{O} \left| R - \frac{R'}{2}  \right> \exp(-\frac{i}{\hbar} R' \cdot P)
\end{equation}
The inverse of the Wigner transform is the Weyl transform $W^{-1}$:
\begin{equation}
    \bra{R} \hat{O}\ket{R'} = \int\frac{dP}{2 \pi \hbar} e^{\frac{i}{\hbar}\left( P \cdot (R-R') \right)} \hat{O}_W\left( \frac{R+R'}{2},P\right)
\end{equation}
Note that the non-commutativity of operators in partially Wignerized space is encoded by the Moyal star product\cite{case:2008:wigner_review}
\begin{equation}
    (\hat{A} \hat{B})_W = \hat{A}_W * \hat{B}_W = \hat{A}_W e^{\frac{i\hbar}{2}\overleftrightarrow{\Lambda}}\hat{B}_W
\end{equation}
where the bidirectional differential operator $\overleftrightarrow{\Lambda}$ is
\begin{equation}
    \overleftrightarrow{\Lambda} = \overleftarrow{\partial}_R \overrightarrow{\partial}_P - \overleftarrow{\partial}_P \overrightarrow{\partial}_R
\end{equation}
The Moyal star product $*$ can be defined as a power series expansion in $\hbar$, known as the Moyal series:
\begin{equation}
    \hat{A}_W * \hat{B}_W = \hat{A}_W\hat{B}_W + \frac{i \hbar}{2} \pb{\hat{A}_W}{\hat{B}_W} + \frac{1}{2}\left( \frac{i \hbar}{2}\right)^2\dpb{\hat{A}_W}{\hat{B}_W} + O(\hbar^3)
\end{equation}
Here $\pb{\hat{A}_W}{\hat{B}_W}$ denotes the Poisson bracket of the two Weyl-symbols, and $\dpb{\hat{A}_W}{\hat{B}_W}$ is the corresponding double-Poisson bracket:
\begin{equation}
    \dpb{\hat{A}_W}{\hat{B}_W} = (\partial_R^2 \hat{A}_W)(\partial_P^2 \hat{B}_W) + (\partial_P^2\hat{A}_W)(\partial_R^2\hat{B}_W)-2(\partial_R\partial_P\hat{A}_W)(\partial_R\partial_P\hat{B}_W)
\end{equation}
This short synopsis should suffice to understand the theory presented below.

In what follows below, our goal will be to derive phase space electronic structure theory and the phase space electronic Hamiltonian in Eq. \ref{eq:PS-intro-hamiltonian} in a novel means using Wigner transforms. Note that, as has already been found in the literature, PSEST goes beyond BH theory insofar as dynamics along a single PSEST surface conserve the total nuclear+electronic momentum whereas dynamics along a single BH surface conserve the total nuclear momentum\cite{tao_basis-free_2025,xuezhi:2023:total_ang_bomd}. Moreover, PSEST naturally recovers a host of physical observables (VCD\cite{duston-VCD-2024,tao-VCD-2024}, Raman optical activity\cite{tao-ROA-2026}, and spin-rotation coupling\cite{peng-SR-2026}) that are not easily recovered with BH theory. Lastly, PSEST recovers the exact hydrogen atom energy spectrum correctly using the reduced electronic mass rather than the raw electronic mass that is found when the nucleus is naively frozen \cite{bian_PS-vibration_2025}. Moreover, Barrera {\em et al.} \cite{polkovnikov:2026:pnas} have shown that  a proper PS treatment of many  non-interacting particles in a  moving box (controlled by a piston) can yield a smooth and intuitive   baseline  around which fluctuations are generated.
Interestingly, below we will show that for nuclear-electronic problems in one dimension, PSEST has one more attribute that has not yet been noted in the literature: namely, the $\hGamma$ operator can have a vanishing non-Abelian curl, which leads to a one-to-one mapping from a PSEST to the exact vibronic Hamiltonian.

The remainder of this work is organized as follows. In Sec. \ref{Sec:BO-Wigner}, we review how to derive the Born-Huang expansion for vibronic Hamiltonians expressed over a finite number of electronic states using  a  Wigner framework. Thereafter, in Sec. \ref{Sec:PS-LF}, we show an analogous expansion of the Hamiltonian can be derived using phase-space  electronic structure framework.  Most interestingly, in Sec. \ref{Sec:PS-CF}, we will show that one can construct a very new formulation for the Hamiltonian in a phase space electronic structure framework provided that $\hGamma$ has zero non-Abelian curl, which is true in one dimension. Turning to applications, in Sec. \ref{Sec:Analytical}, we showcase the analytic power of these approaches with a simple coupled oscillator problem. Finally, Sec \ref{Sec:Numerical} uses these approaches numerically to solve a non-analytic model of hydrogen-bonding. Finally, we summarize and discuss our results in Sec. \ref{Sec:Discussion}, where we address the question of quantum dynamics in two or three dimensions.
 Throughout this work, in order to keep the theory as simple as possible,  we will assume the simplest possible nuclear-electronic Hamiltonian:
\begin{equation}\label{eq:exact-Hamiltonian}
    \hat{H} = \frac{\hat{P}^2}{2M} + \frac{\hat{p}^2}{2m} + \hat{V}(\hat{r},\hat{R}) = \frac{\hP^2}{2M} + \hH_{el}(\hat{r},\hR)
\end{equation}
The most important equations of merit are Eqs. \ref{eq:PS-CF-master-eq}-\ref{eq:newNBH} where we expand the exact, total nuclear-electronic Hamiltonian  within a non-Born-Huang, PSEST framework (and which can be compared against the standard Eq. \ref{eq:BH} for the BH theory). (A second, related PS expansion is also given in Eqs. \ref{eq:PS-LF-master-eq} in the Appendix.)

\section{Review: Born-Oppenheimer and Born-Huang Theory in a Wigner Framework}\label{Sec:BO-Wigner}
Before addressing the main topic of this manuscript (i.e. the derivation of a novel electronic-nuclear Hamiltonian in a  phase space electronic structure framework), let us review how  the traditional Born-Huang approach can be derived through a Wigner framework. To that end, the first step is to perform a partial Wigner transformation of the nuclear operators on Eq. \ref{eq:exact-Hamiltonian}:
\begin{equation}
    \hat{H}_W = \frac{P^2}{2M} + \hat{H}_W^{el}(\hat{r};R)
\end{equation}
Thereafter, we diagonalize the partially-wignerized Hamiltonian at each point in phase space:
\begin{equation}\label{eq:rev-bo-diag}
    \hat{H}_W = \hL_W(R) \hLambda_W(R,P) \hL_W^\dagger(R)
\end{equation}
Eq. \ref{eq:rev-bo-diag} produces a set of Born-Oppenheimer potential energy surfaces (diagonal elements of the diagonal matrix $\hLambda_W$), and Born-Oppenheimer wavefunctions (columns of $\hL_W$). Because $\hH^{el}_W$ is parameterized by only nuclear position ${R}$, $\hL_W$ is therefore only parameterized by nuclear position $R$. As a sidenote, if we treat the lowest potential energy surface $(\hLambda_W)_{00}$ as an effective surface for the nuclei, we obtain the Born-Oppenheimer approximation:
\begin{align}
    \hH^{BO} = W^{-1}\left(\left(\hLambda_W\right)_{00}\right)
\end{align}
In other words, a Weyl transformation acting on $(\hLambda_W)_{00}$ produces an effective Hamiltonian for nuclear vibrations.  

More generally, it  is crucial to note that $\hL_W$ is a unitary transform in nuclear-electronic space. This statement follows because $\hL_W$ is a function of only nuclear position, $\hL_W^\dagger * \hL_W = \hL_W^\dagger \hL_W = \hat{I}$, and thus  $\hL^\dagger \hL= W^{-1}(\hL_W^\dagger*\hL_W) =   \hat{I}$.   
We can therefore obtain an expansion of the original Hamiltonian in terms of our reference Born-Oppenheimer wavefunctions by applying the $\hL$ unitary transformation on the exact Hamiltonian:
\begin{equation}\label{eq:MS-BH-main}
    \begin{aligned}
        \left(\hL^\dagger \hH \hL\right)_W &=  \hL_W^\dagger *\hH_W* \hL_W = \hL_W^\dagger * \left( \hL_W \hLambda_W \hL_W^\dagger\right) * \hL_W\\
        &= \left(\hLambda_W \hL_W^\dagger + \frac{i \hbar}{2} \frac{P}{M} \frac{\partial \hL_W^\dagger}{\partial R} \hL_W \right) * \hL_W   \\
        &= \hLambda_W + i\hbar \frac{P}{M} \frac{\partial\hL_W^\dagger}{\partial R} \hL_W + \hbar^2 \frac{\frac{\partial\hL_W^\dagger}{\partial R} \hL_W \hL_W^\dagger \frac{\partial\hL_W}{\partial R}}{2M} \\ 
        &= \hLambda_W  - i\hbar \frac{P}{M}\hat{D}_R  - \hbar^2 \frac{\hD_R^2}{2M} = \hH_W^{BH}
    \end{aligned}
\end{equation}

The final expression above is the partially Wignerized version of the well-known Born-Huang expansion \cite{born-huang_dynamical_1955} (Eq. \ref{eq:BH}). Note that because $\hLambda_W$ has $P^2$ dependence, the Moyal product expansion terminates at second order. Additionally, we have identified $\hL_W^\dagger \frac{\partial\hL_W}{\partial R} = \hat{D}_R$ as the derivative coupling operator represented in Wigner space. All derivative coupling-like terms arise purely from expansion of the Moyal star products. 

\section{Phase Space Electronic Structure Theory and Littlejohn-Flynn Perturbation Theory}\label{Sec:PS-LF}

In a phase space electronic structure framework, we solve the vibronic problem  starting from   a different electronic ansatz, namely an electronic hamiltonian parametrized by both $R$ and $P$:
\begin{equation}\label{eq:PS-elec-Hamiltonian}
    \hat{H}^{PS}_W (R,P)= \frac{\left(P-i\hbar \hat{\Gamma}\right)^2}{2M} + \hat{H}^{el}_W(R)
\end{equation}
Eq. \ref{eq:PS-elec-Hamiltonian} includes an explicit coupling between nuclear momentum $P$ and the electronic degrees of freedom in the form of an electronic operator $\hat{\Gamma}$, which is designed  to approximate the derivative coupling between adiabatic states. Notably, if such an operator is properly chosen, phase space electronic structure enforces momentum conservation \cite{tao_basis-free_2025}. 
According to PSEST, as suggested in Eq. \ref{eq:rev-bo-diag} above, one diagonalizes Eq. \ref{eq:PS-elec-Hamiltonian} at every phase space point: 
\begin{equation}
    \hat{H}^{PS}_W(R,P) = \hat{L}^{PS}_W \hat{\Lambda}_W^{PS} \hat{L}^{PS\dagger}_W
\end{equation}
This approach produces phase space electronic wavefunctions (the columns of $\hat{L}^{PS}_W$) and phase space surfaces (the diagonal elements of $\hat{\Lambda}_W^{PS}$), both of which now depend on nuclear position $R$ and momentum $P$.

At this point, we note that our research group has already published several articles exploring the nature of the eigenstates and eigenvalues of $\hH^{PS}_W$ in Eq. \ref{eq:PS-elec-Hamiltonian}. 
For instance, if we treat the lowest phase space surface as an effective energy surface for the nuclei $(\hLambda_W^{PS})_{00} = E(R,P)$ (just as in BO theory),  then  Bian {\em et al.}. \cite{bian_PS-vibration_2025} have shown that 
Weyl-transforming this symbol to obtain an effective vibrational Hamiltonian for the nuclei and subsequent
diagonalization 
leads to vibrational energies that are consistently better than the Born-Oppenheimer approximation. 
One can also show that dynamics along single state surfaces roughly recapitulate Nafie's result \cite{nafie:1983:jcp:el_momentum,nafie:1992:vcd,nafie:1997:jpca_current_density1}$\left<\hat{p}_e\right> = m d\left<\hat{r}_e\right>/dt$ (a result that is absent in BO theory)\cite{coraline:2024:jcp:pssh_conserve,tao_basis-free_2025}.
In short, all of our empirical data thus far have clearly demonstrated the benefit of using a single-state PS surface over a single-state BH surface, at least for problems whose vibrations can be reasonably modeled by one electronic state.

Looking ahead, the most interesting question now is if and how one can reconstruct the entire nuclear-electronic Hamiltonian (with many electronic states) starting with a PS electronic structure framework; in other words, can we generate an analog of Eq. \ref{eq:MS-BH-main} outside of a BH framework? In this vein, our only experience so far is through Ref. \cite{Zaidi-electron_transfer-2026}, where we showed empirically that 
for the Shin-Metiu model problem, with crossings between a handful of electronic states, transforming to a pseudo-diabatic representation within a phase space electronic structure framework can  provide a better electronic subspace for describing electron transfer than does Born-Huang theory  \cite{Zaidi-electron_transfer-2026}. Unfortunately, however, beyond numerics, Ref. \cite{Zaidi-electron_transfer-2026} offers little guidance on exactly how to build a diabatic PSEST Hamiltonian in a robust fashion or how to justify that construction.  
 
For a more general theoretical treatment, 
Ref. \cite{littlejohn-flynn_perturbation-theory_1991} by Littlejohn and Flynn (and the subsequent application by Wu {\em et al.} \cite{wu_exact-vibration_2025}) is quite informative. 
Note that a simple expansion shows the connection between the exact (partially Wignerized) Hamiltonian and the PS Hamiltonian:
\begin{equation}
    \begin{aligned}
        \hH^{PS}_W& = \frac{P^2}{2M} + \hH^{el}_W(R) - i\hbar \frac{P}{M}\hGamma - \hbar^2 \frac{\hGamma^2}{2M}\\
        &= \hH_W - i\hbar \frac{P}{M} \hGamma - \hbar^2 \frac{\hGamma^2}{2M}\\
        \implies & \hH_W = \hH^{PS}_W +i\hbar \frac{P}{M}\hGamma + \hbar^2 \frac{\hGamma^2}{2M}
    \end{aligned}
\end{equation}
Thus, in principle one would like to create a phase space representation of the Hamiltonian by conjugation as follows:
\begin{align}\label{eq:isLNormal}
    (\hL^{PS \dagger}\hH\hL^{PS})_W \stackrel{?}{=} \hL_W^{PS\dagger}\star \hH_W \star \hL^{PS}_W
\end{align}

Unfortunately, however, Eq. \ref{eq:isLNormal} is formally invalid because $\hL_W$ is not a unitary transform in nuclear-electronic space, as can be seen from a simple calculation:
\begin{equation}
    \hL_W^{PS \dagger} *\hL^{PS}_W = \hL_W^{PS\dagger} \hL^{PS}_W + \frac{i\hbar}{2}\pb{\hL^{PS\dagger}_W}{\hL^{PS}_W} + \cdots \neq \hat{I} 
\end{equation}
The inequality arises above because PS wavefunctions depend on both $R$ and $P$, so that the poisson bracket $\pb{\hL_W^{PS\dagger}}{\hL^{PS}_W}\neq0$. Thus $\hL^{PS}_W$ is not unitary in nuclear-electronic space.

The inequality above was recognized long ago by  Littlejohn and Flynn \cite{littlejohn-flynn_perturbation-theory_1991}, who proposed that one can use perturbation theory to find a nearby unitary matrix that is star unitary.  
Mathematically, we suppose that $\hL_W$ is sufficiently close to star unitarity $\left( \text{i.e.} \pb{\hL_W^{PS\dagger}}{\hL^{PS}_W} \text{is sufficiently small} \right)$ that we can design a perturbatively corrected $\hL'_W$ that satisfies star unitarity:
\begin{equation}\label{eq:L-exp}
    \hL_W' = \hL^{PS}_W + \hbar\hL_{1} + \hbar^2 \hL_2 + \cdots
\end{equation}
One then enforces star unitarity order by order to obtain a closed form expression for $\hL_1$, $\hL_2$, etc.
\begin{equation}\label{eq:L'-unitarity}
    \hL_W'^\dagger * \hL_W' = \hat{I}
\end{equation}
Thereafter, using the fact that $\hL'$ is unitary (in the full nuclear-electronic space), one can transform the total vibronic Hamiltonian in Wigner space:
\begin{equation}\label{eq:PS-LF-main}
    \hLambda_W'  = \hL_W'^\dagger * \left( \hH^{PS}_W +i\hbar \frac{P}{M}\hGamma + \hbar^2 \frac{\hGamma^2}{2M} \right) *\hL_W'
    = \hLambda_W + \hbar \hLambda_1 + \hbar^2 \hLambda_2 + \cdots
\end{equation}
Although $\hLambda_W$ is a diagonal operator, the operators $\hLambda_j$ for $j \ge 1$ connect different phase space surfaces together, just as in BH theory. Unlike BH theory, however,   
the expansion  in Eq. \ref{eq:PS-LF-main} does not necessarily terminate at 2nd order. Moreover, because the expansion relies on the Moyal series (just as the derivative couplings in BH theory), there is no guarantee that these corrections are perturbatively small \cite{heller-wignerPS-1976}. 

Given this caveat, we will now work out the first and second order wavefunction corrections. From the unitary conditions in Eqs. \ref{eq:L-exp} and \ref{eq:L'-unitarity}, the first order corrections have the form:
\begin{equation}\label{eq:L-1st}
    \hat{L}_1 = \hL^{PS}_W\left(-\frac{i}{4}\pb{\hL_W^{PS\dagger}}{\hL^{PS}_W} + \hat{A} \right)
\end{equation}
Here, $\hat{A}$ is an anti-hermitian matrix representing a gauge freedom in our choice of $\hat{L}_1$. Similarly, the second order wavefunction corrections take the form:
\begin{equation}\label{eq:Y-2nd}
    \hat{L}_2 = \hL_W \left( -\frac{1}{2} \hL^\dagger_1 \hL_1 - \frac{i}{4}\left( \pb{\hL^\dagger_1}{\hL^{PS}_W} + \pb{\hL_W^{PS\dagger}}{\hat{L}_1} \right) + \frac{1}{16}\dpb{\hat{L}_W^{PS\dagger}}{\hL^{PS}_W} + \hat{B}\right)
\end{equation}
Again, there is an anti-hermitian matrix representing a gauge freedom ($B$). 

In this work, because our goal is to investigate the expansion in Eq. \ref{eq:PS-LF-main} rather than any single state phase space electronic structure surfaces, we will choose the gauge where $\hat{A}=\hat{B}=0$. Note that this distinguishes our multi-state expansion from traditional Littlejohn-Flynn perturbation theory, according to which one chooses  $\hat{A},\hat{B}$ by diagonalizing the total Hamiltonian electronic block by electronic block. The latter approach was studied by Wu {\em et al.} \cite{wu_exact-vibration_2025}, where they choose $\hat{A}$ such that corrections lie solely on the diagonal of $\hLambda_W'$ before Weyl transforming the lowest perturbatively corrected electronic surface. That being said, 
$\hat{A}$ and $\hat{B}$ can become large near avoided crossings so such an approach would not be feasible for a stable multi-state approach; by contrast, setting $\hat{A} = \hat{B} = 0$ is always stable.
Thus, we will label the above multi-state phase space approach based on Littlejohn-Flynn  theory as PS-LF.
For the final, exact expressions for the first and second order Hamiltonians, $\hLambda_1^{PS-LF}$ and $\hLambda_2^{PS-LF}$, see Appendix \ref{app:PS-LF-Lambda}.

\section{Phase Space Electronic Hamiltonian through the Non-Abelian Curvature Condition}\label{Sec:PS-CF}

The approach detailed in the section above (and in Appendix \ref{app:PS-LF-Lambda}) for PSEST is rigorous and always applicable. That being said, under certain circumstances, $\hGamma$ takes on a special form. In particular, it is well known\cite{mead:1992:rmp} that if the non-Abelian curl of $\hGamma$ is zero, 
\begin{align}\label{eq:ch4-nonabelian}
    \hat{\Omega}^{A\alpha,B\beta} \equiv \frac{\partial \hat{\Gamma}_{B,\beta}}{\partial R_{A,\alpha}} - \frac{\partial \hat{\Gamma}_{A,\alpha}}{\partial R_{B,\beta}} - \left[ \hat{\Gamma}_{A,\alpha},\hat{\Gamma}_{B,\beta} \right] \stackrel{?}{=} 0
\end{align}
then one can write
\begin{equation}\label{eq:gamma-CC}
    \hat{\Gamma}_{A,\alpha} = \frac{\partial \hat{U}^\dagger_W}{\partial R_{A,\alpha}} \hat{U}_W = - \hU_W^\dagger \frac{\partial \hU_W}{\partial R_{A,\alpha}}
\end{equation}
for some unitary matrix $\hat{U}(R)$ that depends only on nuclear position $R$.

Eq. \ref{eq:ch4-nonabelian} is the   powerful curl condition\cite{mbaer:1975:cpl,mead:1992:rmp,littlejohn:2022:jcp:parallel} and Eq. \ref{eq:gamma-CC} allows for a very clear interpretation of the nuclear-electronic Hamiltonian through the lens of PSEST. Note that, as shown in the Discussion section, Eq. \ref{eq:gamma-CC} does indeed hold for our particular choices of the $\hGamma$ operator whenever the dynamics occur in one dimension.

Given that Eq. \ref{eq:gamma-CC} holds, we can obtain a PS representation for the exact vibronic Hamiltonian as follows. First let us similarity transform the total Hamiltonian with a nuclear-electronic unitary transformation $\hU$ which does not depend on $\hat{P}$:
\begin{equation}\label{eq:H-tilde}
    \hat{\tilde{H}} = \hat{U} \hat{H} \hat{U}^\dagger = \frac{\left(\hat{P}-i\hbar \hU \frac{\partial \hU^\dagger}{\partial R}\right)^2}{2M} + \hat{U} \hat{H}^{el}(\hat{r},\hat{R}) \hat{U}^\dagger 
\end{equation}
Second, we perform a partial Wigner transformation (Wignerizing only the nuclear degrees of freedom):
\begin{align}
    \hat{\tilde{H}}_W = \left( \hat{U} \hat{H} \hat{U}^\dagger \right)_W &= \frac{\left(P- i\hbar \left(\hat{U} \frac{\partial \hat{U}^\dagger}{\partial R}\right)_W\right)^2}{2M} + \left(\hat{U} \hat{H}^{el}(\hat{r},\hat{R}) \hat{U}^\dagger\right)_W \label{eq:easy1} \\
    &= \frac{\left(P-i\hbar \hat{U}_W \frac{\partial \hat{U}^\dagger_W}{\partial R}\right)^2}{2M} + \hat{U}_W \hat{H}^{el}_W(\hat{r};R)\hat{U}^\dagger_W \label{eq:easy2}
\end{align}
To derive Eq. \ref{eq:easy2} from Eq. \ref{eq:easy1}, we have used the fact that $\hat{U}_W$, $\partial \hat{U}^\dagger_W/\partial R$ and $\hat{H}_W^{el}$ all depend only on $R$ (and not on $P$) so that star products are equivalent to normal multiplication.

At this point, we have introduced a momentum dependent coupling of the form $-i\hbar \hU_W\partial_R\hU_W^\dagger \cdot P/M$, but our Hamiltonian is still in the similarity transformed frame. Thus, third, to return to the original basis, we  reverse the similarity transform and conjugate by $\hU_W(R)$ at every  point $(R,P)$ in phase space: 
\begin{equation}\label{eq:PS-from-CC}
    \hH^{PS}_W \stackrel{def}{=} \hU_W^\dagger\hat{\tilde{H}}_W\hU_W = \hat{U}_W^\dagger \left( \hat{U} \hat{H}\hat{U}^\dagger\right)_W \hat{U}_W = \frac{\left( P - i\hbar \frac{\partial \hat{U}_W^\dagger}{\partial R} \hat{U}_W \right)^2}{2M} + \hat{H}^{el}_W(\hat{r};R)
\end{equation}

In Eq. \ref{eq:PS-from-CC} above, we have derived the PSEST Hamiltonian cited in Eq. \ref{eq:PS-elec-Hamiltonian}. Most importantly, reversing the steps taken above, we find that 
\begin{equation}\label{eq:exact-from-PS}
    \hat{H} = \hat{U}^\dagger  W^{-1}\left( \hat{U}_W \hat{H}^{PS}_W \hat{U}_W^\dagger \right) \hat{U}
\end{equation}
Thus, there is no information lost when working with a PS electronic structure framework.  In fact, provided that Eq. \ref{eq:ch4-nonabelian} holds, PSEST can be viewed merely as   ``preconditioner'' to make the BH expansion more efficient. After all, as far as BH theory is concerned, the equivalent of Eq. \ref{eq:exact-from-PS} above for the  Hamiltonian is the tautological expression:
\begin{align}
    \hH = W^{-1}\left(\hH_W\right)
\end{align}
which is equivalent to setting $\hU = \hat{I}$ in Eq. \ref{eq:exact-from-PS}.

Finally, the theory above dictates that one can generate the exact electronic-nuclear Hamiltonian
by working with $\hat{\tilde{H}}$ instead of $\hH$. Here, let us write:
\begin{equation}\label{eq:h-tilde-def}
    \left( \hU \hH \hU^\dagger \right)_W = \hat{\tilde{H}}_W = \hat{U}_W \hat{H}^{PS}_W \hat{U}_W^\dagger = \hat{U}_W \hat{L}^{PS}_W \hat{\Lambda}^{PS}_W \hat{L}_W^{PS\dagger} \hat{U}_W^\dagger \equiv \hat{Y}_W \hLambda^{PS}_W \hat{Y}_W^\dagger
\end{equation}
where we have defined $\hY_W = \hU_W\hL^{PS}_W$. Just as above (for standard Littlejohn-Flynn theory), in order to 
develop a power series expansion in $\hbar$, we must begin by  perturbing the phase-space wavefunctions $\hY_W$ to star unitarity:
\begin{align}
    \hY'_W = \hY_W + \hbar \hY_1 &+ \hbar^2 \hY_2 + \cdots\\
    \hY'^\dagger_W * \hY'_W &= \hat{I}
\end{align}
Thereafter, we can use $\hY'$ to directly transform $\hat{\tilde{H}}$:
\begin{equation}\label{eq:PS-CF-main}
    \begin{aligned}
        \hat{\Lambda}'^{PS-CF} &= \hY_W'^\dagger * \hat{\tilde{H}}_W * \hY_W' = \hat{Y}_W'^\dagger * (\hat{U}_W \hat{H}^{PS}_W \hat{U}_W^\dagger)*\hat{Y}_W' = \hY_W'^\dagger *(\hY_W \hLambda_W \hY_W^\dagger)*\hY_W' \\
        &= \hLambda + \hbar \hLambda_1 + \hbar^2 \hLambda_2 + \cdots
    \end{aligned}
\end{equation}

Because such an expansion relies on having zero non-Abelian curl for $\hGamma$, we will label this multi-state curl-free phase space expansion as PS-CF. As far  as imposing star unitarity, we find that the first and second order wavefunction corrections for $Y$ takes 
the same form for PS-CF as they do for $\hL_W$ for PS-LF. The first order correction is:
\begin{equation}\label{eq:Y-1st}
    \hat{Y}_1 = \hat{Y}_W\left(-\frac{i}{4}\pb{\hat{Y}_W^\dagger}{\hat{Y}_W} + \hat{A} \right)
\end{equation}
The  2nd order correction is:
\begin{equation}\label{eq:Y-2nd}
    \hat{Y}_{2} = \hat{Y}_W \left( -\frac{1}{2} \hat{Y}^\dagger_1 \hat{Y}_1 - \frac{i}{4}\left( \pb{\hat{Y}^\dagger_1}{\hat{Y}_W} + \pb{\hat{Y}_W^\dagger}{\hat{Y}_1} \right) + \frac{1}{16}\dpb{\hat{Y}_W^\dagger}{\hat{Y}_W} + \hat{B}\right)
\end{equation}
Again, we introduce antihermitian matrices $\hat{A},\hat{B}$ representing a gauge freedom and will choose the gauge where $\hat{A}=\hat{B}=0$.

Next, let us evaluate the corrections to $\hLambda_W$. While these expressions utilize Poisson brackets, they can be made to resemble standard BH theory in an intuitive fashion by using the following identities:
\begin{gather}
    \hY_W^\dagger \frac{\partial \hY_W}{\partial_R} = \hat{D}_R,\quad \hY_W^\dagger \frac{\partial \hY_W}{\partial P} = \hat{D}_P\\
    \hat{D}_R^\dagger = -\hat{D}_R,\quad \hat{D}_P^\dagger = -\hat{D}_P\\
    \implies \hY_1 = \hY_W\left(-\frac{i}{4}\pb{\hY_W^\dagger}{\hY_W}\right) = \hY_W \left(\frac{i}{4} \com{\hD_R}{\hD_P}\right)
\end{gather}
Here, we use $\com{\hat{A}}{\hat{B}},\acom{\hat{A}}{\hat{B}}$ to denote the commutator and anti-commutator respectively. Expanding the right hand side of Eq. \ref{eq:PS-CF-main} to 2nd order in $\hbar$ and performing some algebra yields the 1st and 2nd order corrections to $\hLambda_W$:
\begin{equation}
    \begin{aligned}
        \hat{\Lambda}_1^{PS-CF} =  -&\hY_W^\dagger \hY_1 \hLambda_W - \hLambda_W \hY_1^{\dagger}\hY_W - \frac{i}{2}\pb{\hLambda_W}{\hY_W^\dagger}\hY_W - \frac{i}{2}\hY_W^\dagger \pb{\hY_W}{\hLambda_W} 
        \\&-\frac{i}{2}\left( \hY_W^\dagger \frac{\partial \hY_W}{\partial R} \hLambda_W \frac{\partial \hY_W^\dagger}{\partial P} \hY_W - \hY_W^\dagger\frac{\partial \hY_W}{\partial P} \hLambda_W \frac{\partial \hY_W^\dagger}{\partial R} \hY_W\right)
        \end{aligned}
\end{equation}
Plugging in the definition of $\hY_1$ further yields: 
\begin{equation}
    \begin{aligned}
        \hLambda_1^{PS-CF} =& -\frac{i}{2}\pb{\hLambda_W}{\hY_W^\dagger}\hY_W -\frac{i}{2} \hY_W^\dagger\pb{\hY_W}{\hLambda_W} + \frac{i}{4}\pb{\hY_W^\dagger}{\hY_W}\hLambda_W + \frac{i}{4}\hLambda_W\pb{\hY_W^\dagger}{\hY_W} 
        \\&-\frac{i}{2}\left( \hY_W^\dagger \frac{\partial \hY_W}{\partial R} \hLambda_W \frac{\partial \hY_W^\dagger}{\partial P} \hY_W - \hY_W^\dagger\frac{\partial \hY_W}{\partial P} \hLambda_W \frac{\partial \hY_W^\dagger}{\partial R} \hY_W\right)\\
        =& -\frac{i}{2}\left( \hD_R \frac{\partial \hLambda_W }{\partial P} + \frac{\partial \hLambda_W }{\partial P} \hD_R -  \hD_P \frac{\partial \hLambda_W }{\partial R} - \frac{\partial \hLambda_W}{\partial R} \hD_P \right)  
        \\ & + \frac{i}{2}\left( \hD_R \hLambda_W \hD_P - \hD_P \hLambda_W \hD_R \right)
        \\ & - \frac{i}{4}\left( \com{\hD_R}{\hD_P}\hLambda_W + \hLambda_W \com{\hD_R}{\hD_P}\right) 
    \end{aligned}
\end{equation}
Above, we find that the derivative couplings that appear in BH theory generalize within PSEST to include $\hD_P$ contributions. 
Similarly, the second order energy correction is:
\begin{equation}\label{eq:lambda-2nd}
    \begin{aligned}
        \hLambda_2^{PS-CF} =& \hY_2^\dagger \hY_W \hLambda_W + \hY^\dagger_1 \hY_W \hLambda_W \hY_W^\dagger \hY_1 + \hLambda_W  \hY_W^\dagger \hY_2 \\
        &+ \frac{i}{2} \left(\hY_W^\dagger\pb{\hY_W \hLambda_W \hY_W^\dagger}{\hY_1} + \hY^\dagger_1\pb{\hY_W \hLambda_w \hY_W^\dagger}{\hY_W} + \pb{\hY^\dagger_1}{\hY_W \hLambda_W } + \pb{\hY_W^\dagger}{\hY_W \hLambda_W \hY_W^\dagger \hY_1}\right)\\
        &-\frac{1}{8} \left( \hY_W^\dagger \dpb{\hY_W \hLambda_W \hY_W^\dagger}{\hY_W} + 2 \pb{\hY_W^\dagger}{\pb{\hY_W \hLambda_W \hY_W^\dagger}{\hY_W}} + \dpb{\hY_W^\dagger}{\hY_W \hLambda_W } \right)
    \end{aligned}
\end{equation}
Unlike the 1st order correction, attempting to expand the second order correction in terms of derivative couplings in $R$ and $P$ will lead to an explosion of terms, so we will leave this expression in Poisson bracket form for the sake of clarity. For a more detailed expression, see Appendix \ref{App:2ndOrderExpansion}.

Finally, let us separate the terms that appear in Born-Huang theory $\hH^{BH}_W$ from non-Born-Huang terms unique to a phase space theory (i.e. terms which require $\partial_P \hY \neq 0$). To second order, we find:
\begin{align}
        \label{eq:PS-CF-master-eq}
        \hH^{PS-CF(2)}_W &= \hH^{BH}_W + \hH_{W
        }^{NBH} \\
        \label{eq:newBH}
        \hH^{BH}_W & = \hLambda_W -\frac{i\hbar}{2} \left(D_R \hLambda_P + \hLambda_P D_R\right) - \frac{\hbar^2}{4} D_R \hLambda_{PP} D_R -\frac{\hbar^2}{8} \acom{\hLambda_{PP}}{D_R^2}\\
        \label{eq:newNBH}
        \hH^{NBH}_W & = \frac{i\hbar}{2} \left( -\hD_P \hLambda_R - \hLambda_R\hD_P + \hD_R\hLambda_W\hD_P - \hD_P\hLambda_W\hD_R\right)
        \\
        &  -\frac{i\hbar}{4} \left(\com{\hD_R}{\hD_P}\hLambda_W + \hLambda_W\com{\hD_R}{\hD_P}\right) \nonumber
        \\
        & +\hbar^2 \hLambda_2^{PS-CF} + \left(\frac{\hbar^2}{4} \hD_R \hLambda_{PP} \hD_R +\frac{\hbar^2}{8} \acom{\hLambda_{PP}}{\hD_R^2}\right) + \ldots \nonumber
 \end{align}
Here, we define $\partial_R \hLambda_W = \hLambda_R $, $ \partial_P \hLambda_W =\hLambda_P$, $\partial^2_P \hLambda_W =\hLambda_{PP}$, etc.  for notational simplicity. Note that Eq. \ref{eq:newBH} is equivalent to the standard Born-Huang equation in Eq. \ref{eq:MS-BH-main}, but only when $\hLambda_P= P/M$. The bottom line is that a PS electronic structure Hamiltonian resembles the BH Hamiltonian in many ways, but with some important differences, the most obvious one being the existence of derivative couplings in both R ($D_R$) {\em and} in P ($D_P$).
As a side note, in the last line above, for the sake of cleanliness, we have chosen not to expand out the 2nd order term and simply subtract out the terms which lead to the 2nd order Born-Huang correction. A full derivation of Eq. \ref{eq:newBH} from \ref{eq:lambda-2nd} can be found in Appendix \ref{app:2ndOrderBH}.

\begin{figure}[H]
    \centering
    \includegraphics[width=1.0\linewidth]{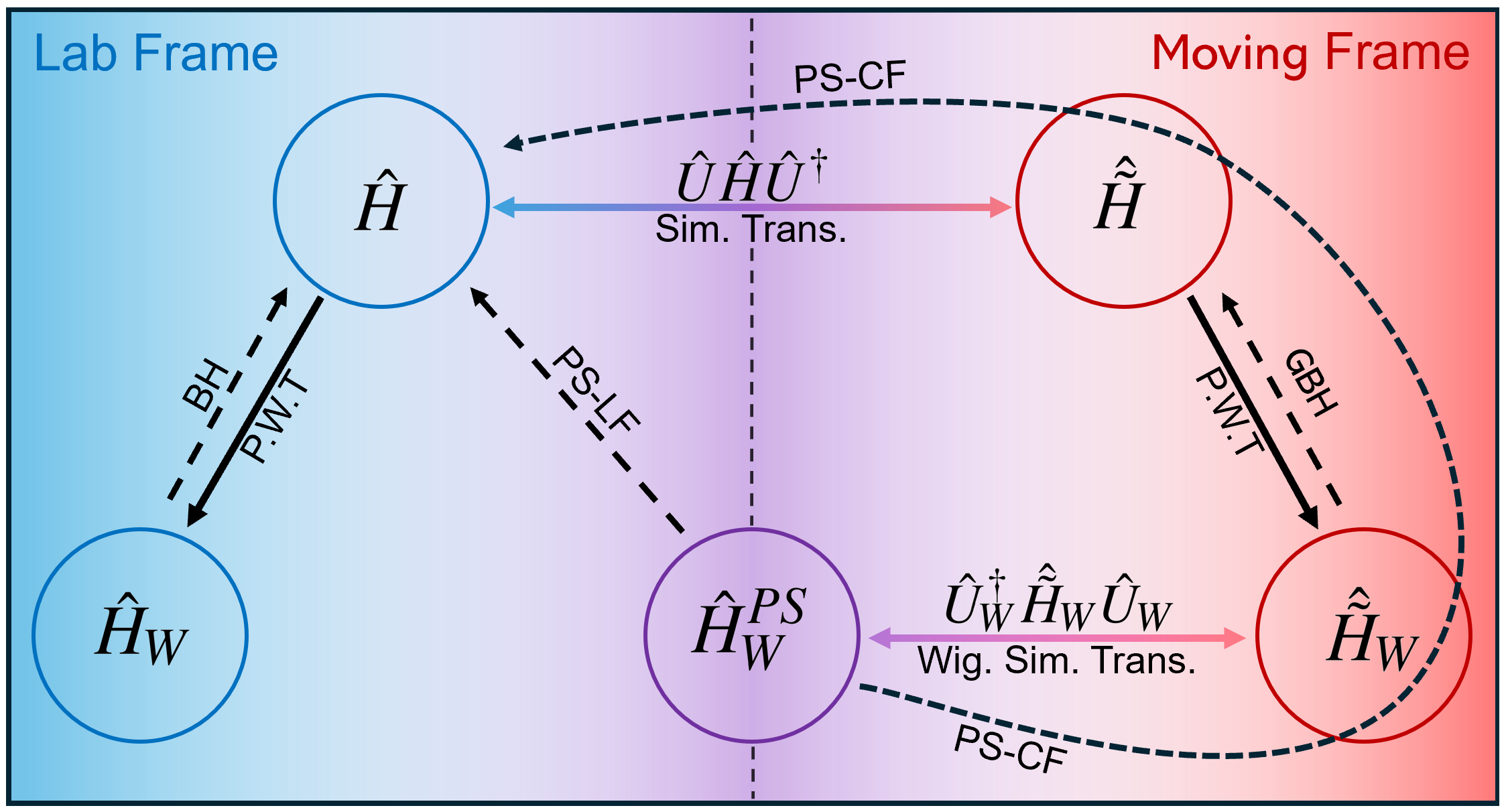}
    \caption{
    A diagrammatic summary of the theory presented above. On the left, we draw a conceptual figure of standard BH theory. If we start with the full, quantum Hamiltonian $\hat{H}$, a partial Wigner transform  yields $\hat{H}_W$. If  one diagonalizes $\hH_W$ and generates adiabatic electronic states, the resulting partially Wignerized matrix representation of the Hamiltonian can be inverted by a Weyl transformation to return to the exact Hamiltonian; this famous Born-Huang expansion terminates at second order. On the right hand side of the figure, in a moving frame, we note that one can begin with the standard Hamiltonian $\hH$, similarity transform with $\hU$ to $\hat{\tilde{H}}$, perform a Wigner transform, and then perform another similarity transform with $\hU^{\dagger}_W$ to generate $\hH_W^{PS}$. If $\hH_W^{PS}$ is further diagonalized, then  the resulting matrix representation of the Hamiltonian can also be inverted to recover $\hH$ -- the only difference being that the expansion (PS-LF or PS-CF) is of infinite order. In this paper, the PS-LF equations are given in Appendix \ref{app:PS-LF-Lambda} and the PS-CF equations are given in Eqs. \ref{eq:PS-CF-master-eq}-\ref{eq:newNBH} above. Finally, note that we can consider the Weyl transform of $\hat{\tilde{H}}_W$ to $\hat{\tilde{H}}$ to be a generalized Born-Huang  (GBH) expansion -- which again is an infinite (not quadratic) order expansion. }
    \label{fig:mapping}
\end{figure}

A summary of the theory above is provided in Fig. \ref{fig:mapping}. Note that, because we work with Wigner-Weyl transform -- which are more general than the simple notion of freezing nuclei within Born Huang -- the language might appear unfamiliar. For instance, the Born-Huang expansion can be thought of as the inverse of a partial Wigner transform of the standard Hamiltonian $\hH_W$. More generally, one can also define a generalized Born-Huang expansion as the inverse of the partial Wigner transform of the Hamiltonian in a moving frame. By mixing and matching Wigner and similarity/unitary transforms, many different representations are possible.

\section{An Analytical Result: Coupled Quantum Harmonic Oscillators}\label{Sec:Analytical}
The theoretical models developed above will be examined using a coupled quantum harmonic oscillator problem that can be solved analytically. This approach allows us to dissect how exactly Born-Huang, PS-LF, and PS-CF differ in their representation of the Hamiltonian. (After the  initial publication of this article on the arXiv, we became aware that another analysis of various electronic structure approaches with exactly solvable problems has also been developed recently by Wiggins.\cite{wiggins_2026_solvable_models_reveal_born_oppenheimer})
Consider the following 2-particle 1D Hamiltonian:
\begin{equation}\label{eq:anal-H}
    \hH = \frac{\hat{P}^2}{2M} + \frac{\hat{p}^2}{2m} + \frac{K}{2} \hat{R}^2 + \frac{g}{2}(\hat{r} - \hat{R})^2
\end{equation}
A light particle $m$ is harmonically coupled to a heavy particle $M$ with spring constant $g$, while the heavy particle is harmonically bound to the origin with spring constant $K$. We will consider the case $M> m$ for some fixed spring constants $K,g$. For simplicity, we will call the light and heavy particles the electron and nucleus respectively.
Below we will compare the lowest vibronic gap and the zero point energy for the exact case versus with Born-Huang theory, as well as PS-LF and PS-CF expansions. These results are plotted in Fig. \ref{fig:Analytic-PT2}. In Appendix \ref{app:response}, we will further analyze the low frequency behavior of the nuclear response function as well as the latter's poles.

\subsection{Exact Solution}\label{subSec:Anal-exact}
Eq. \ref{eq:anal-H} can be diagonalized exactly through a normal mode decomposition. 
The Hessian for this system is:
\begin{equation}
    \mathcal{H} 
    =
    \begin{bmatrix}
        g & -g \\
        -g & K+g
    \end{bmatrix}
\end{equation}
Thus, the mass-weighted hessian is:
\begin{equation}
    F = \mathcal{M}^{-1/2} \mathcal{H} \mathcal{M}^{-1/2} 
    =
    \begin{bmatrix}
        \frac{g}{m} & \frac{-g}{\sqrt{mM}} \\
        \frac{-g}{\sqrt{mM}} & \frac{K+ g}{M}
    \end{bmatrix}
\end{equation}
The eigenvalues of this matrix correspond to the square of the decoupled oscillator frequencies. Some algebra leads to the following result for the decoupled oscillator frequencies:

\begin{equation}\label{eq:ex-eigenfreqs}
    \omega_\pm = \sqrt{\frac{gM + (K+g)m \pm \sqrt{(gM-(K+g)m)^2 + 4g^2mM}}{2mM}}
\end{equation}
The eigenspectrum is
\begin{equation}
    E_{n_+,n_-} = \hbar \omega_+\left( n_+ + \frac{1}{2} \right) + \hbar\omega_- \left( n_- + \frac{1}{2} \right)
\end{equation}
The lowest vibronic gap is therefore
\begin{equation}\label{eq:Analytic-Exact-Gap}
    \Delta E^{exact}= \hbar \omega_-
\end{equation}
and the zero point energy for the total ground state is:
\begin{equation}\label{eq:Analytic-Exact-ZPE}
    E_{ZPE}^{exact} = \frac{\hbar}{2}(\omega_+ + \omega_-)
\end{equation}

We perform a power-series expansion for the lowest vibronic gap with respect to the mass ratio 
\begin{align}
\label{eq:mM_eta} \eta=m/M
\end{align}
The exact solution expanded to 2nd order in $\eta$ is:
\begin{equation}\label{eq:analytic-exact-gap}
    \Delta E^{exact}  = 
    \hbar\sqrt{\frac{K}{M}}\left(1 - \frac{\eta}{2} + \left(\frac{3}{8} - \frac{K}{2g}\right)\eta^2 + O(\eta^3)  \right)
\end{equation}
We will use this as a reference to compare our approximate methods. Note that there are two terms that arise at second order in $\eta$. To understand how each second order term arises, consider the limit when $g/K \rightarrow \infty$:

\begin{align}
    \lim_{g/K \rightarrow \infty} \Delta E^{exact} &= \hbar \sqrt{\frac{K}{M+m}}  = \hbar \sqrt{\frac{K}{M}} \left( 1 - \frac{\eta}{2} + \frac{3}{8}\eta^2 + \cdots \right)
\end{align}
Only $(3/8) \eta^2$ is present in this limit. Physically, this corresponds to the electron and nucleus being bound by a rigid rod, co-moving as a single effective oscillator with mass $M+m$. At finite $g/K$, the electron does not follow the nucleus exactly, producing an additional frequency-dependent inertial correction. Therefore, we will denote $(3/8)\eta^2$ and $-(K/2g)\eta^2$ as the ``static" and ``dynamic" second order term respectively.  
\subsection{Born-Huang Theory}\label{subSec:Anal-BH}
\subsubsection{Single State Approximation}
As described above, we start by considering the partially Wignerized Hamiltonian
\begin{align}
    \hH_W(R,P) = \frac{P^2}{2M } +  \frac{\hat{p}^2}{2m} + \frac{K}{2}\hat{R}^2 + \frac{g}{2}(\hat{r}-R)^2 
\end{align}
and  obtain adiabatic electronic wavefunctions and eigensurfaces: 
\begin{equation}
    \begin{aligned}
        &\hat{H}_{W}(R,P) \ket{\psi_n^{BH}(R)} = \Lambda_{nn}(R,P) \ket{\psi_n^{BH}(R)}\\
        &\psi_n^{BH}(r;R) = \phi_n(r-R)\\
        &\Lambda_{nn}(R,P) = \frac{P^2}{2M} +  \frac{K}{2}R^2 + \hbar \sqrt{\frac{g}{m}} \left( n + \frac{1}{2}\right)
    \end{aligned}
\end{equation}
We denote $\phi_n$ as the $n$th quantum harmonic oscillator wavefunction.

Truncating this expression to the lowest electronic state (i.e. $n=0$), re-quantizing, and diagonalizing the resulting effective vibrational Hamiltonian leads to the Born-Oppenheimer result:
\begin{equation}
    E^{BO}_\nu = \hbar \sqrt{\frac{K}{M}} \left( \nu + \frac{1}{2} \right) + \frac{\hbar}{2}\sqrt{\frac{g}{m}} = \hbar \Omega\left( \nu + \frac{1}{2} \right) + \frac{\hbar\omega}{2} 
\end{equation}
Therefore, we obtain a Born-Oppenheimer lowest vibronic gap:
\begin{equation}\label{eq:Analytic-BO-Gap}
    \Delta E^{BO} = \hbar \Omega =\hbar \sqrt{\frac{K}{M}}
\end{equation}
and zero point energy:
\begin{equation}\label{eq:Analytic-BO-ZPE}
    E^{BO}_{ZPE} = \frac{\hbar}{2}(\Omega + \omega) =\frac{\hbar}{2} \left(\sqrt{\frac{K}{M}} + \sqrt{\frac{g}{m}} \right)
\end{equation}

An asymptotic expansion of the Born-Oppenheimer gap (Eq. \ref{eq:Analytic-BO-Gap}) is accurate to only order $O(1)$ relative to Eq. \ref{eq:analytic-exact-gap} and has an error that scales as $O(\eta)$. This finding is visualized in Fig. \ref{fig:Analytic-PT2}; the Born-Oppenheimer (BO) result consistently scales the worst (order $O(\eta)$) and has the highest error when compared to the other tested approaches. Notably, the Born-Oppenheimer gap corresponds to the well-known adiabatic limit\cite{born:huang}:

\begin{equation}
    \lim_{M/m\rightarrow \infty} \Delta E^{exact} = \hbar \sqrt{ \frac{K}{M} } = \Delta E^{BO}
\end{equation}

In this limit, the electron mass is negligible, so the lowest vibronic gap is determined solely by the nuclear mass.

\subsubsection{Multi-State Approximation to 1st Order}
Next, let us include the first order Born-Huang correction when the first excited electronic state is included. Because $\partial_R\psi_n^{BH}=-\partial_r\psi_n^{BH} $, we find $\hD_R=\hat{p}/i\hbar$. 
Including the first order Born-Huang correction yields:
\begin{equation}
    \hH^{BH(1)}_W(R,P) = \frac{P^2}{2M} + \frac{K}{2} R^2 + \hbar \sqrt{\frac{g}{m}}\left(\hat{n} + \frac{1}{2}\right) -\frac{P}{M}\hat{p} 
\end{equation}

To continue analytically, we truncate to two electronic states and re-quantize $R,P$ via Weyl transform. The resulting spin-boson model is
\begin{equation}
    \hH^{BH(1)} = \left( \frac{\hat{P}^2}{2M} + \frac{K}{2}\hat{R}^2 + \hbar\omega \right) \hat{\sigma}_0 
    - \frac{\hbar \omega}{2} \hat{\sigma}_z 
    - \hbar \alpha_p\frac{\hat{P}}{M}\hat{\sigma}_y\\
\end{equation}
Here, we define $\alpha_p=\sqrt{\frac{m\omega}{2\hbar}}$ and nuclear bosonic ladder operators $\hat{b}^\dagger,\hat{b}$. 
Since $M \gg m$ and the 1st order coupling term scales as $\sqrt{m/M}$, we can approximate the final term with Rayleigh-Schrodinger perturbation theory (RSPT) to second order, $\hH^{BH(1)} = \hH_0 + \hat{V}$, where
\begin{align}
    \hH_0 &= \left( \frac{\hat{P}^2}{2M} + \frac{K}{2}\hat{R}^2 + \hbar\omega \right) \hat{\sigma}_0 
    - \frac{\hbar \omega}{2}\hat{\sigma}_z \\
    \hat{V} &= - \hbar \alpha_p\frac{\hat{P}}{M}\hat{\sigma}_y
    = -i\frac{\hbar}{2} \sqrt{\frac{m}{M}}\sqrt{\omega \Omega}(\hat{b}^\dagger - \hat{b})\hat{\sigma}_y
\end{align}
We can write $\hH_0$ in terms of the nuclear bosonic number operator $\hat{N}$:
\begin{align}
\hat{N} &= \hat{b}^{\dagger}\hat{b}\\
    \hH_0 &= \left(\hbar\Omega\left(\hat{N} + \frac{1}{2}\right) + \hbar \omega \right) \hat{\sigma}_0 - \frac{\hbar \omega}{2}\hat{\sigma}_z
\end{align}

It is now straightforward to evaluate the 2nd order perturbative energy corrections to the lowest two eigenstates of $\hat{H_0}$, $\ket{0,0}$ and $\ket{1,0}$ (indexed as $\ket{N,n}$): 
\begin{equation}
    \begin{aligned}
        E_{0,0}^{(2)} &= \frac{|\bra{0,0}\hat{V}\ket{1,1}|^2}{E_{0,0} - E_{1,1}} 
        = \frac{1}{-\hbar(\omega + \Omega)} \left(\frac{\hbar}{2} \sqrt{\frac{m}{M}\omega\Omega}\right)^2 
        = -\frac{\hbar}{4} \frac{m}{M}\frac{ \omega \Omega}{\omega + \Omega}\\
        E_{1,0}^{(2)} &=  \frac{|\bra{1,0}\hat{V}\ket{0,1}|^2}{E_{1,0} - E_{0,1}} + \frac{|\bra{1,0}\hat{V}\ket{2,1}|^2}{E_{1,0} - E_{2,1}} = \frac{\hbar}{4} \frac{m}{M} \omega \Omega \left( \frac{1}{\Omega - \omega} -  \frac{2}{\Omega + \omega} \right)
    \end{aligned}
\end{equation}
Thus, we predict a  gap of the form:
\begin{align}\label{eq:Analytic-BH-Gap}
    \Delta E^{BH(1)} \approx \hbar\Omega +  E^{(2)}_{1,0} - E^{(2)}_{0,0} = \hbar\Omega \left( 1 - \frac{m}{2M} \frac{\omega^2}{\omega^2 -\Omega^2} \right)
\end{align} 
and zero point energy:
\begin{equation}\label{eq:Analytic-BH-ZPE}
    E_{ZPE}^{BH(1)} \approx \frac{\hbar}{2} (\omega + \Omega) - \frac{\hbar}{4} \frac{m}{M} \frac{\omega \Omega}{\omega + \Omega} = \frac{\hbar}{2} \left( \omega + \Omega - \frac{m}{2M} \frac{\omega \Omega}{\omega + \Omega} \right)
\end{equation}

\subsubsection{Multi-State Approximation to 2nd Order} 
We start with the full (second order) Born-Huang expansion of the Hamiltonian: 

\begin{equation}
    \hH = \frac{P^2}{2M} + \frac{K}{2} R^2 + \hbar \sqrt{\frac{g}{m}}\left(\hat{n} + \frac{1}{2}\right) -\frac{P}{M}\hat{p} + \frac{\hat{p}^2}{2M}
\end{equation}
Truncation to two states yields:
\begin{equation}
    \hH^{BH(2)} = \left( \frac{\hat{P}^2}{2M} + \frac{K}{2}\hat{R}^2 + \hbar\omega\left(1 + \frac{m}{2M}\right)\right)\hat{\sigma}_0 
    - \frac{\hbar \omega}{2}\left(1 + \frac{m}{2M}\right)\hat{\sigma}_z 
    - \hbar \alpha_p\frac{\hat{P}}{M}\hat{\sigma}_y
\end{equation}
Including the 2nd order correction renormalizes the electronic gap and zero-point electronic energy by the mass ratio $\frac{m}{2M}$. Therefore, we obtain an expression for the lowest vibronic gap and zero point energy similar to what was found at first order:
\begin{equation}\label{eq:Analytic-BH2-Gap}
    \Delta E^{BH(2)} = \hbar \Omega\left( 1  - \frac{m}{2M} \left(1 + \frac{m}{2M}\right)\frac{\omega^2}{\omega^2(1+m/2M)^2 - \Omega^2} \right)
\end{equation}

\begin{equation}\label{eq:Analytic-BH2-ZPE}
    E_{ZPE}^{BH(2)} = \frac{\hbar}{2}\left( \Omega + \omega \left( 1 + \frac{m}{2M} \right) - \frac{m}{2M} \frac{\omega \Omega}{\omega(1+m/2M) + \Omega} \right) 
\end{equation}

Asymptotic expansion of the 1st and 2nd order corrected Born-Huang gaps (Eqs. \ref{eq:Analytic-BH-Gap} \& \ref{eq:Analytic-BH2-Gap}) yields:
\begin{align}
    \Delta E ^{BH(1)} &= \hbar \sqrt{\frac{K}{M}} \left( 1 - \frac{\eta}{2} - \frac{K}{2g}\eta^2 + O(\eta^3)  \right)\\
    \Delta E^{BH(2)} &= \hbar \sqrt{\frac{K}{M}} \left( 1 - \frac{\eta}{2} - \left(\frac{1}{4} + \frac{K}{2g}\right)\eta^2 + O(\eta^3) \right)
\end{align}

Both Born-Huang approaches have an error which scales as $O(\eta^2)$. Including the first order correction exactly reproduces the ``dynamic" second order term found in the exact calculation while BH(2) does not accurately reproduce the ``static" second order term. This is clearly evident when taking the limit $g/K \rightarrow \infty$:
\begin{align}
    \lim_{g/K\rightarrow \infty} \Delta E^{BH(1)} &= \hbar \sqrt{\frac{K}{M}} \left(1 - \frac{m}{2M}\right) = \hbar \sqrt{\frac{K}{M}}\left(1 - \frac{\eta}{2}\right)\\
    \lim_{g/K \rightarrow \infty} \Delta E^{BH(2)} &= \hbar \sqrt{\frac{K}{M}}\left( \frac{1}{1 + \frac{m}{2M}} \right) = \hbar\sqrt{\frac{K}{M}}\left( 1 - \frac{\eta}{2} - \frac{1}{4}\eta^2 + \cdots \right)
\end{align}

This is because the Born-Huang expansion uses BO wavefunctions which, for this Hamiltonian, are fully real valued and therefore have zero mean electronic momentum. If one wanted to obtain the exact mass renormalization from the light particle, it would require including {\em all} electronic states in the calculation. As seen in Fig. \ref{fig:Analytic-PT2}, adding the first order Born-Huang correction (BH(1)) improves the scaling as well as the absolute error, but including the 2nd order correction gives a diminishing improvement while maintaining the same error scaling.

\subsection{Phase Space Electronic Structure Theory}\label{subSec:Anal-PS}

\subsubsection{Single State Approximation
}
Let us now address the same problem within a phase space electronic structure framework. 
For this  problem, we set $\hGamma = \hat{p}/i\hbar$, yielding a phase space electronic Hamiltonian of the form: 

\begin{equation}
    \hH^{PS}_W = \frac{(P - \hat{p})^2}{2M} + \frac{\hat{p}^2}{2m} + \frac{g}{2}(\hat{r} - R)^2 + \frac{K}{2}\hat{R}^2
\end{equation}
Diagonalizing this Hamiltonian at each $(R,P)$ yields PS wavefunctions and energies:
\begin{equation}
    \begin{aligned}
        \hat{H}^{PS}_W (R,P) \ket{\psi_n^{PS}(R,P)}= \Lambda_{nn}(R,P) \ket{\psi_n^{PS}(R,P)} \\
        \Lambda_{nn}(R,P) = \frac{P^2}{2(M+m)} + \frac{1}{2}KR^2 + \hbar \sqrt{\frac{g}{\mu}}\left(n + \frac{1}{2}\right)\\
        \psi^{PS}_n(r;R,P) = \exp\left(\frac{i}{\hbar}\frac{\mu}{M}P \cdot r\right)\phi_n(r-R)
    \end{aligned}
\end{equation}
Here, we have defined 
\begin{align}
\mu = \frac{mM}{m+M} \label{eq:defmu}    
\end{align}
Just as in Born-Oppenheimer theory, one can restrict dynamics to the lowest eigensurface and construct a single-state vibrational Hamiltonian:
\begin{equation}
    \hH^{SS-PS} = W^{-1}(\Lambda_0) = \frac{\hat{P}^2}{2(M+m)}  +\frac{K}{2}\hat{R}^2  + \frac{\hbar}{2}\sqrt{\frac{g}{\mu}}
\end{equation}
Solving this effective Hamiltonian yields an eigenspectrum that is slightly different from the Born-Oppenheimer result in Sec. \ref{subSec:Anal-BH} above:
\begin{equation}
    E_\nu^{SS-PS} = \hbar\sqrt{\frac{K}{m+M}} \left( \nu + \frac{1}{2} \right) +\frac{\hbar}{2}\sqrt{\frac{g}{\mu}}
\end{equation}
Thus, the lowest vibronic gap and zero point energy are:
\begin{equation}\label{eq:Analytic-PS-Gap} 
    \Delta E^{SS-PS} = \hbar \sqrt{\frac{K}{m+M}}
\end{equation}
\begin{equation}\label{eq:Analytic-PS-ZPE}
    E_{ZPE}^{SS-PS} = \frac{\hbar}{2} \left( \sqrt{\frac{K}{m+M}} + \sqrt{\frac{g}{\mu}} \right)
\end{equation}

Now, consider the asymptotically expanded single-state phase space vibronic gap (Eq. \ref{eq:Analytic-PS-Gap}):
\begin{equation}
    \begin{aligned}
        \Omega^{SS-PS}&=\hbar\sqrt{\frac{K}{M}}\left(1 - \frac{\eta}{2} + \frac{3}{8} \eta^2 + O(\eta^3) \right)
    \end{aligned}
\end{equation}
We see that the single-state phase space result has the same order of error as both BH(1) and BH(2). While BH theory captures the ``dynamic" second order term, SS-PS obtains the ``static" second order term. In the adiabatic limit $\omega \gg \Omega$, SS-PS would predict a more accurate gap and would scale better than BO. Encouragingly, according to Fig. \ref{fig:Analytic-PT2} the single-state PS approach consistently outperforms BH(1) and is systematically more accurate than BH(2) with increasing force constant ratio $g/K$. To understand why SS-PS performs so well, consider the limits $g/K \rightarrow \infty$:
\begin{align}
     \lim_{g/K \rightarrow \infty} \Delta E^{SS-PS} &= \hbar \sqrt{\frac{K}{M+m}} = \lim_{g/K \rightarrow \infty}\Delta E^{exact}
\end{align}
We find that SS-PS corresponds to a \textit{different} 
limit than BO wherein the heavy and light particle are always co-moving and therefore act as one  particle with effective mass $M+m$.
If one further takes the limit that $M/m \rightarrow \infty$, one recovers the BO result. Since SS-PS theory does not couple electronic states, the effective vibrational Hamiltonian does not have information on the excited electronic state and therefore cannot obtain the ``dynamic" second order term.

\subsubsection{PS-LF Expansion}
In order to evaluate the first order PS-LF correction, we first evaluate the derivative couplings with respect to both $R$ and $P$. A simple calculation yields $\hD_R = \hat{p}/i\hbar$, $\hD_P = \frac{i}{\hbar}\frac{\mu}{M} \left( R \hat{I} + \hat{r}\right)$ and $\com{\hD_R}{\hD_P} = -\frac{i}{\hbar}\frac{\mu}{M}\hat{I}$. The first order correction is therefore:
\begin{equation}
        \hLambda_1^{PS-LF(1)}(R,P) = \frac{\mu}{M} \left(\frac{P^2}{M} - KR^2\right)\hat{I} + \frac{\hat{p}^2}{2M} + \frac{\mu}{M^2}P\hat{p} + \frac{\mu}{M} \left(\frac{g}{2}-K \right)R \hat{r} + \frac{\mu}{M}\frac{g}{2}\hat{r}^2
\end{equation}
We then truncate to two electronic states:
\begin{equation}
    \begin{aligned}
        H^{PS-LF(1)}_W = & \begin{bmatrix}
            H_0 + \frac{1}{2}(1 +\frac{\mu}{M})\hbar\tilde{\omega} 
            & \frac{\mu}{M}\left[\left(\frac{g}{2} - K\right)\tilde{r}R-i\frac{\tilde{p}}{M} P\right]\\
            \frac{\mu}{M}\left[\left(\frac{g}{2} - K\right)\tilde{r}R + i\frac{\tilde{p}}{M} P\right] 
            & H_0 + \frac{3}{2}(1 +\frac{\mu}{M})\hbar\tilde{\omega}
        \end{bmatrix}\\
        H_0 =& \frac{P^2}{2(m+M)} + \frac{K}{2}R^2 + \frac{\mu}{M} \left(\frac{P^2}{M} - KR^2\right)\\
        =& \frac{P^2}{2} \left(\frac{1}{m+M} + \frac{2\mu}{M^2} \right) + \frac{K(1-\frac{2\mu}{M})}{2}R^2
    \end{aligned}
\end{equation}
For a compact description,  we have defined the following electronic operators and variables 
\begin{gather}
    \tilde{\omega} = \sqrt{\frac{g}{\mu}}\\
    \hat{r} = \tilde{r}  \left(\hat{a}^\dagger + \hat{a}\right)  \\
    \tilde{r} = \sqrt{\frac{\hbar}{2\mu\tilde{\omega}}}\\
    \hat{p} =  i\tilde{p}  \left(\hat{a}^\dagger-\hat{a}\right) \\
    \tilde{p} = \sqrt{\frac{\hbar\mu\tilde{\omega}}{2}} 
\end{gather}
and nuclear harmonic oscillator operators and variables:

\begin{gather}
    \tilde{\Omega} = \sqrt{\frac{\tilde{K}}{\tilde{M}}} = \sqrt{\frac{K(1-2\mu/M)}{\frac{M(M+m)}{M+2m}}} = \sqrt{\frac{K}{M}}\sqrt{\frac{(M+2m)(M-m)}{(M+m)^2}}\\
    \hat{R} = \sqrt{\frac{\hbar}{2\tilde{M}\tilde{\Omega}}} \left(\hat{b}^\dagger + \hat{b}\right) = \tilde{R}\left(\hat{b}^\dagger + \hat{b}\right)\\
    \hat{P} = i\sqrt{\frac{\hbar\tilde{M}\tilde{\Omega}}{2}} \left(\hat{b}^\dagger - \hat{b}\right) = i\tilde{P}\left(\hat{b}^\dagger - \hat{b}\right)
\end{gather}

We then re-quantize the heavy particle degrees of freedom $R,P$ via Weyl transform and obtain the spin-boson model:
\begin{gather}
    \hat{H}^{PS-LF(1)} = \begin{bmatrix}
        \hbar \tilde{\Omega} \left( \hat{N} + \frac{1}{2}\right) + \frac{\hbar \tilde{\omega}}{2}\left(1 + \frac{\mu}{M}\right)
        &(G_R + G_P)\hat{b}^\dagger + (G_R-G_P)\hat{b}
        \\
        (G_R - G_P)\hat{b}^\dagger + (G_R + G_P)\hat{b}
        & \hbar \tilde{\Omega} \left( \hat{N} + \frac{1}{2}\right) + \frac{3\hbar \tilde{\omega}}{2}\left(1 + \frac{\mu}{M}\right)
    \end{bmatrix}\\
    G_R = \frac{\mu}{M} \left( \frac{g}{2} - K \right)\tilde{r}\tilde{R}, \qquad 
    G_P = \frac{\mu}{M^2} \tilde{p}\tilde{P}
\end{gather}
As before, we perform 2nd order RSPT to obtain a correction for the lowest vibronic states:
\begin{equation}
    \begin{aligned}
        E_{0,0}^{(2)} &= -\frac{1}{\hbar}\frac{|G_R-G_P|^2}{\tilde{\omega}(1 + \mu/M) + \tilde{\Omega}}\\
        E_{1,0}^{(2)}&=-\frac{1}{\hbar}\left( \frac{2|G_R-G_P|^2}{\tilde{\omega}(1 + \mu/M) + \tilde{\Omega}} + \frac{|G_R + G_P|^2}{\tilde{\omega}(1 + \mu/M) - \tilde{\Omega}}\right)\\
         E_{1,0}^{(2)} -  E_{0,0}^{(2)}  &=-\frac{1}{\hbar}\left( \frac{|G_R-G_P|^2}{\tilde{\omega}(1 + \mu/M) + \tilde{\Omega}} + \frac{|G_R + G_P|^2}{\tilde{\omega}(1 + \mu/M) - \tilde{\Omega}}\right)
    \end{aligned}
\end{equation}
Thus, in the end, including first order corrections and only two electronic states, we predict the following gap and zero point energy;    \begin{align}\label{eq:Analytic-MSLF-Gap}
        \Delta E^{PS-LF(1)} &= \hbar \tilde \Omega - \frac{2\tilde{\omega}(G_R^2 + G_P^2) + \tilde{\Omega}G_RG_P}{\hbar(\tilde{\omega}^2 - \tilde{\Omega}^2)}\\
        &=\hbar \tilde{\Omega} - \frac{\hbar \mu}{2 M^2 \tilde{M}\tilde{\omega}\tilde{\Omega}} \frac{\tilde{\omega}\left(1+\frac{\mu}{M}\right)\left( \left( \frac{g}{2} - K \right)^2 + \left( \frac{mM}{2m+M}\tilde{\omega}{\tilde{\Omega}} \right)^2 \right) + 2\tilde{\Omega}\left( \frac{g}{2} - K \right)\left( \frac{mM}{2m+M}\tilde{\omega}{\tilde{\Omega}} \right)}{\tilde{\omega}^2(1+\mu/M)^2 - \tilde{\Omega}^2} \nonumber\\
        \label{eq:Analytic-PSLF-ZPE}
        E_{ZPE}^{PS-LF(1)} &= \frac{\hbar}{2}\left(\tilde{\Omega} + \tilde{\omega}\left( 1 + \frac{\mu}{M} \right)\right)  - \frac{(G_R - G_P)^2}{\hbar (\tilde{\omega}(1+\mu/M) + \tilde{\Omega})}\\
        &= \frac{\hbar}{2}\left( \tilde{\Omega} + \tilde{\omega}\left(1 + \frac{\mu}{M}\right) - \frac{\mu}{2M^2\tilde{M}\tilde{\omega}\tilde{\Omega}} \frac{\left( \frac{g}{2} - K - \frac{mM}{2m+M}\tilde{\omega}\tilde{\Omega} \right)^2}{\tilde{\omega}(1+\mu/M) + \tilde{\Omega}} \right) \nonumber
\end{align}

Now, we asymptotically expand the PS-LF(1) gap (Eq. \ref{eq:Analytic-MSLF-Gap})
\begin{equation}
    \Delta E^{PS-LF(1)} = \hbar \sqrt{\frac{K}{M}} \left( 1 - \frac{\eta}{2} - \left(\frac{1}{8} + \frac{K}{2g} + \frac{g}{8K}\right)\eta^2 + O(\eta^3) \right)  
\end{equation}
The result is still accurate to 1st order in $\eta$ and obtains the correct $K/g$ term. However, it worsens the static renormalization and includes an erroneous $g/K$ term not present in the exact solution.
This ramification arises because PS-LF approach targets $\hH$ rather than $\hat{\tilde{H}}$, so it ``undoes" some of the mass-renormalization caused by $\hGamma$ in the process of targeting $\hH$. In fact, according to  Fig. \ref {fig:Analytic-PT2}, the vibronic gap predicted by PS-LF(1) theory is worse than all Born-Huang approaches and sometimes even BO theory when $g/K \gg 1$ and $M/m \approx 10$.

\subsubsection{PS-CF Expansion}
Because $\hGamma = \hat{p}/i\hbar$ does not depend on $R$, it has zero non-Abelian curl and Eq. \ref{eq:gamma-CC} holds:

\begin{equation}\label{eq:Anal-U}
    \hU = \exp\left( \frac{i}{\hbar} \hat{R}\hat{p} \right)
\end{equation}
Eq. \ref{eq:Anal-U} corresponds to rotating into the moving frame with respect to the heavy coordinate $R$. 
A similarity transformation by $\hU$ and a partial Wigner transformation of  the Hamiltonian yields:
\begin{equation}
    \hat{\tilde{H}}_W = \frac{(P - \hat{p})^2}{2M} + \frac{\hat{p}^2}{2m} + \frac{g}{2}\hat{r}^2 + \frac{K}{2}\hat{R}^2
\end{equation}
Thereafter, we obtain $\hH_W^{PS}$ through point-wise reversal of the similarity transformation (Eq. \ref{eq:h-tilde-def}). Since we seek a multi-state expansion with respect to the solutions of $\hH^{PS}_W$, we must also apply the reverse unitary transformation to the PS wavefunctions. 

\begin{equation}\label{eq:Analytic-PS-CF-wfs}
    \psi_n^{'PS}(R,P) = e^{\frac{i}{\hbar}\frac{\mu}{M}P \cdot r}\phi_n(r)
\end{equation}

As with Born-Oppenheimer theory, these wavefunctions have a global phase freedom as a function of the phase-space variables $(R,P)$. We choose to align the phases in Eq. \ref{eq:Analytic-PS-CF-wfs} such that $\bra{n(R,P)} \partial_R \ket{n(R,P} = 0 $ and $\bra{n(R,P)} \partial_P \ket{n(R,P} = 0$. 

The first order correction is then simple:
\begin{equation}
    \hLambda_1^{PS-CF}(R,P) = -\frac{\mu}{M}KR\hat{r}
\end{equation}
As above, we truncate to two electronic states and requantize via Weyl transform to obtain a spin-boson model:
\begin{equation}
    \hat{H}^{PS-CF(1)} = \begin{bmatrix}
        \hbar\tilde{\Omega}\left(\hat{N} + \frac{1}{2}\right) + \hbar\tilde{\omega}/2 
        & -\hbar\gamma \left( \hat{b}^\dagger + \hat{b}\right) \\
        -\hbar\gamma \left( \hat{b}^\dagger + \hat{b}\right)
        & \hbar\tilde{\Omega}\left(\hat{N} + \frac{1}{2}\right) + 3\hbar\tilde{\omega}/2
    \end{bmatrix}
\end{equation}
We define $\tilde{\Omega}^2 = \frac{K}{m + M}$ and $\gamma = \frac{\mu}{M}K\sqrt{\frac{1}{2\mu\tilde{\omega}} }\sqrt{\frac{1}{2(m+M)\tilde{\Omega}}}$ for brevity. Again as above, we perform 2nd order Rayleigh-Schrodinger perturbation theory to obtain a closed form expression for the correction to the gap. The result is:
\begin{equation}
    \begin{aligned}
        \delta E^{(2)}_{0,0} &= \frac{|\bra{0,0}\hat{V}\ket{1,1}|^2}{E_{0,0} - E_{1,1}} =-\hbar\frac{\gamma^2}{\tilde{\omega} + \tilde{\Omega}}\\
        \delta E^{(2)}_{1,0} &= \frac{|\bra{1,0}\hat{V}\ket{0,1}|^2}{E_{1,0} - E_{0,1}} + \frac{|\bra{1,0}\hat{V}\ket{2,1}|^2}{E_{1,0} - E_{2,1}} = -\hbar\gamma^2 \left( \frac{1}{\tilde{\omega} - \tilde{\Omega}} + \frac{2}{\tilde{\omega} + \tilde{\Omega}}\right)
    \end{aligned}
\end{equation}

This leads to the lowest vibronic gap and zero point energy:
\begin{equation}\label{eq:Analytic-MSCC-Gap}
    \Delta E^{PS-CF(1)} = \hbar \tilde{\Omega}\left( 1 - \frac{m}{2M} \frac{\tilde{\Omega}^2}{\tilde{\omega}^2 - \tilde{\Omega}^2}\right)
\end{equation}

\begin{equation}\label{eq:Analytic-MSCC-ZPE}
    E_{ZPE}^{PS-CF(1)} = \frac{\hbar}{2}\left( \tilde{\omega} + \tilde{\Omega} - \frac{m}{2M} \frac{\tilde{\Omega}^3}{\tilde{\omega}(\tilde{\omega} + \tilde{\Omega})} \right)
\end{equation}

\begin{figure}
    \centering
    \includegraphics[width=1\linewidth]{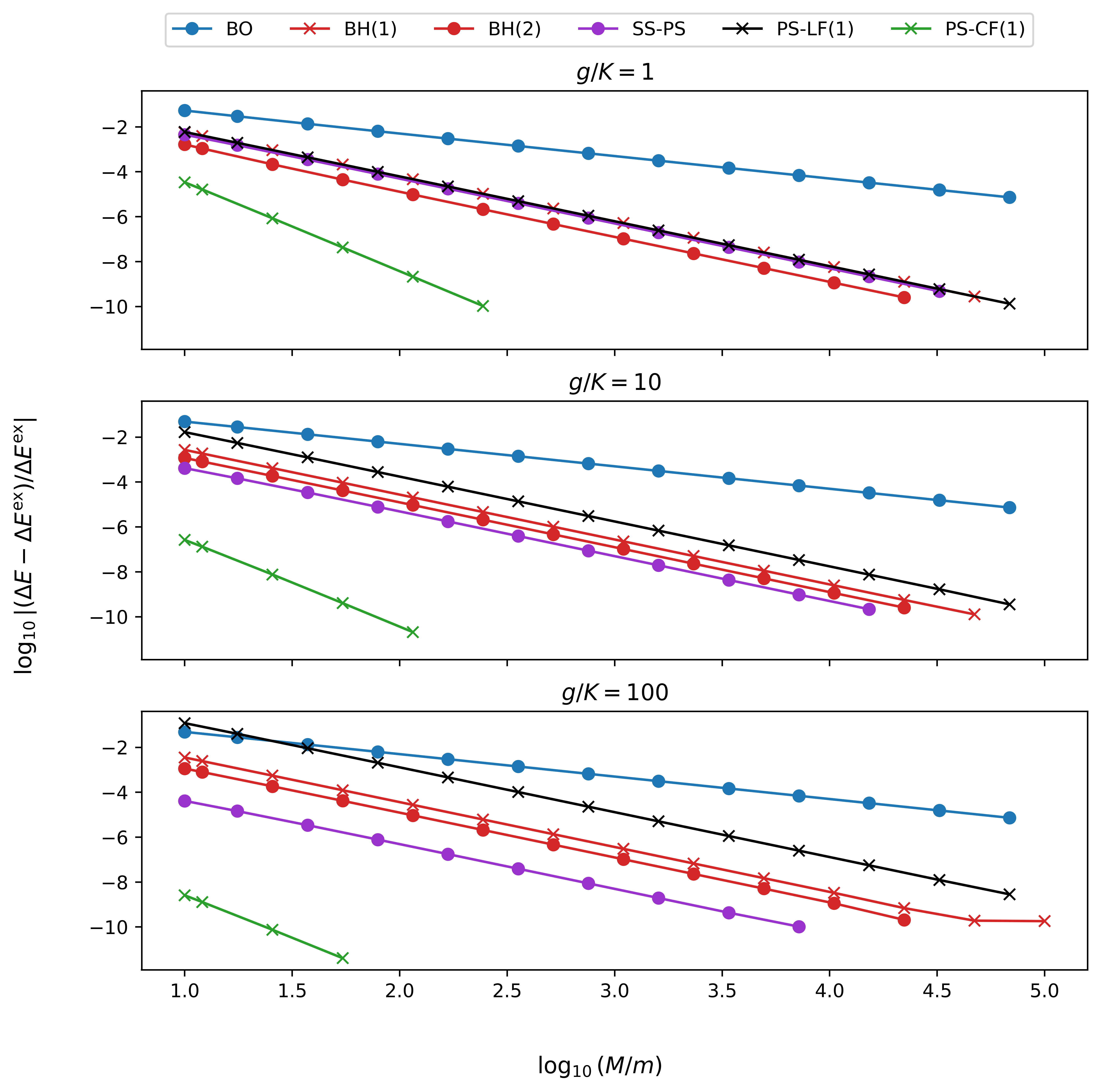}
    \caption{  Relative error of the lowest vibronic energy gap on a log scale as a function of mass ratio at various force constant ratios $g/K$ for the coupled oscillator problem. Here, we plot the Born-Oppenheimer result (Eq. \ref{eq:Analytic-BO-Gap}), as well as the first order corrected (Eq. \ref{eq:Analytic-BH-Gap}) and second order corrected (Eq. \ref{eq:Analytic-BH2-Gap}) Born-Huang vibronic gap.
    We also plot SS-PS (Eq. \ref{eq:Analytic-PS-Gap}), PS-LF(1) (Eq. \ref{eq:Analytic-MSLF-Gap}) and PS-CF(1) (Eq. \ref{eq:Analytic-MSCC-Gap}) vibronic gaps. We find that PS-CF(1) calculations consistently outperform and scale better than all other tested methods, while PS-LF(1) actually leads to a worse result than SS-PS calculations. (Higher order corrections are needed for PS-LF to yield strong results  -- see results in Fig. \ref{fig:borgis} for the Borgis problem.) Plots are truncated when absolute error is smaller than numerical precision.}
    \label{fig:Analytic-PT2}
\end{figure}

Finally, we expand the 1st order PS-CF gap (Eq. \ref{eq:Analytic-MSCC-Gap}):
\begin{equation}
    \Delta E^{PS-CF(1)} = \hbar \sqrt{\frac{K}{M}}\left(1 - \frac{\eta}{2} + \left(\frac{3}{8} - \frac{K}{2g} \right)\eta^2 + O(\eta^3) \right)
\end{equation}
PS-CF(1) accurately captures the ``dynamic" second order term, thus matching the exact expansion up to 2nd order in $\eta$. Most notably, Fig. \ref{fig:Analytic-PT2} indicates the PS-CF(1) result consistently outperforms every other method and has an error that scales as $\sim O(\eta^{3.8-4.0})$.

\subsection{Zero Point Energies}

Finally, let us address the question of the asymptotic convergence of zero point energies.
Since zero point energies depend on $\sqrt \eta$, we define $\zeta = \sqrt{\eta} = \sqrt{m/M}$ and expand with respect to $\zeta$. For notational brevity, we define $\omega = \sqrt{g/m}$, and $\omega_K = \sqrt{K/m}$. The exact zero point energy (Eq. \ref{eq:Analytic-Exact-ZPE}) is expanded as:
\begin{align}
    E_{ZPE}^{exact} &= \frac{\hbar}{2}\left[\sqrt{\frac{g}{m}}\left(1 + \frac{\zeta^2}{2} + \left(\frac{K}{2g} - \frac{1}{8}\right) \zeta^4 + \cdots \right) + \sqrt{\frac{K}{m}}\left(\zeta - \frac{\zeta^3}{2} + \left( \frac{3}{8} - \frac{K}{2g}\right)\zeta^5 +\cdots \right)\right]\\
    &= \frac{\hbar}{2} \left( \omega + \omega_K\zeta + \frac{\omega}{2} \zeta^2 - \frac{\omega_K}{2}\zeta^3 + \omega\left(\frac{\omega_K^2}{2\omega^2} - \frac{1}{8}\right)\zeta^4 + \omega_K\left(\frac{3}{8} - \frac{\omega_K^2}{2\omega^2}\right)\zeta^5 + O(\zeta^6)  \right)
\end{align}

The ZPE expansions for the approximate methods tested are:
\begin{align}
    E_{ZPE}^{BO} &= \frac{\hbar}{2} \left( \omega + \omega_K\zeta \right)\\
    E_{ZPE}^{BH(1)} &= \frac{\hbar}{2} \left( \omega + {\omega_K \zeta} - \frac{\omega_K}{2}\zeta^3 + \frac{\omega_K^2}{2\omega}\zeta^4 - \frac{\omega_K^3}{2\omega^2}\zeta^5+ O(\zeta^6) \right)\\
    E_{ZPE}^{BH(2)} &= \frac{\hbar}{2} \left( \omega + \omega_K\zeta  + \frac{\omega}{2}\zeta^2 - t{\frac{\omega_K}{2}\zeta^3} + \frac{\omega^2_K}{2\omega}\zeta^4  + \omega_K \left( \frac{1}{4} - \frac{\omega_K^2}{2\omega^2} \right)\zeta^5+  O(\zeta^6) \right)\\
    E_{ZPE}^{SS-PS} &= \frac{\hbar}{2} \left( \omega + \omega_K \zeta + \frac{\omega}{2} \zeta^2 - t{\frac{\omega_K}{2}\zeta^3} - \frac{\omega}{8} \zeta^4 + \frac{3\omega_K}{8}\zeta^5 + O(\zeta^6) \right)\\
    E_{ZPE}^{PS-LF(1)} &= \frac{\hbar}{2} \left( \omega + {\omega_K \zeta} + \frac{3\omega}{2} \zeta^2 - \frac{\omega_K}{2} \zeta^3 - \frac{5\omega}{8}\zeta^4 - \omega_K\left( \frac{\omega_K^2}{2\omega^2} + \frac{1}{8} - \frac{\omega^2}{8\omega_K^2} \right)\zeta^5 + O(\zeta^6) \right)\\
    E_{ZPE}^{PS-CF(1)} &= \frac{\hbar}{2} \left( \omega + \omega_K \zeta + \frac{\omega}{2} \zeta^2 - t{\frac{\omega_K}{2}\zeta^3} - \frac{\omega}{8} \zeta^4 + \omega_K\left(\frac{3}{8} - \frac{\omega_K^2}{2\omega^2}\right)\zeta^5 + O(\zeta^6) \right)
\end{align}

Asymptotic expansions reveal important higher order differences between the approximate methods tested even when leading error scalings are similar. The BO approximation simply truncates the expression to first order. Similarly, BH(1) omits the $O(\zeta^2)$ term but recovers the correct $O(\zeta^3)$ term and higher order ``dynamic" terms. BH(2) recovers the correct $O(\zeta^2)$ and $O(\zeta^3)$ terms, but lacks the fourth order ``static" term and underestimates the fifth order ``static" term. PS-LF(1) performs less favorably: in addition to reproducing the $O(\zeta)$ term, it gives an incorrect $O(\zeta^2)$ static contribution, and therefore differs from the exact expansion already at second order. Finally, PS-CF(1) reproduces the exact expansion through $O(\zeta^3)$ and also recovers the exact fifth-order dynamic contribution, but misses the fourth-order term. This behavior is analogous to BH(1), suggesting that a higher-order PS-CF correction could potentially improve the asymptotic accuracy in the same way that going from BH(1) to BH(2) restores the missing second-order contribution.

\begin{figure}
    \centering
    \includegraphics[width=1\linewidth]{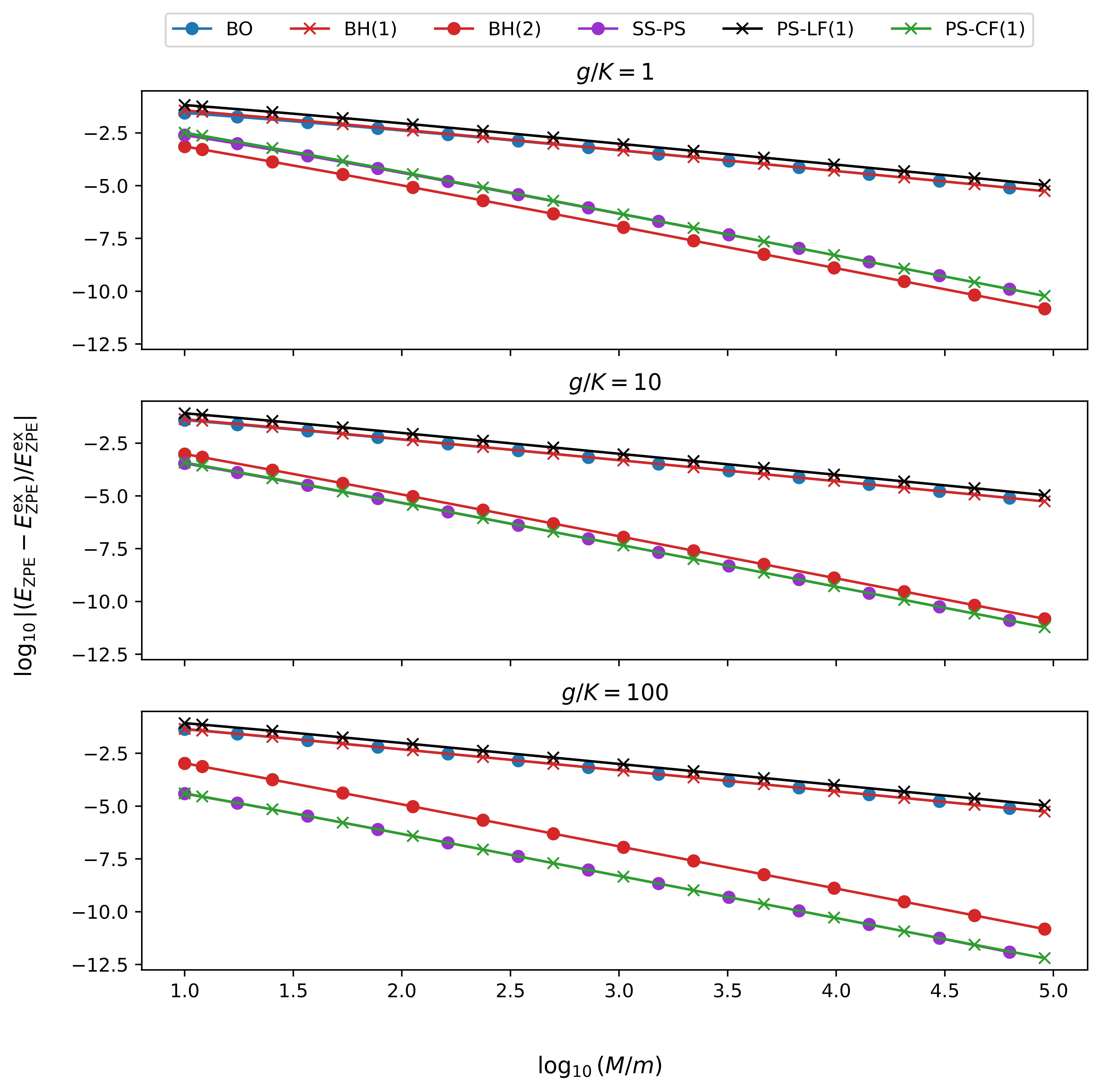}
    \caption{Relative error of the zero point energy on a log scale as a function of mass ratio at various force constant ratios $g/K$ for the coupled oscillator problem. Here, we plot the Born-Oppenheimer results (Eq. \ref{eq:Analytic-BO-ZPE}), as well as the first order corrected (Eq. \ref{eq:Analytic-BH-ZPE}) and second order corrected (Eq. \ref{eq:Analytic-BH2-ZPE}) Born-Huang results. We also plot single-state phase space (Eq. \ref{eq:Analytic-PS-ZPE}), PS-LF(1) (Eq. \ref{eq:Analytic-PSLF-ZPE}) and PS-CF(1) (Eq. \ref{eq:Analytic-MSCC-ZPE}) zero point energies. We find that, at large $g/K$, SS-PS consistently outperforms BH(2). For this problem, PS-LF(1) is consistently the worst estimate; one simply must go to higher order for a meaningful correction.}
    \label{fig:Analytic-ZPE-PT2}
\end{figure}

The analysis above for the predicted zero point energies is plotted in Fig. \ref{fig:Analytic-ZPE-PT2}. Perhaps most notably, in the limit $g/K \gg 1$, we find that SS-PS consistently outperforms BH(2) while scaling at the same rate. 

More generally, adding PS corrections to first order does not lead to any improvement in the zero point energies. In fact, PS-LF(1) performs worse than BO while scaling at the same rate. Higher order corrections are necessary -- e.g. see results for the Borgis problem in the next section.  More generally, we have found that the vibrational gap is  a simpler and more straightforward metric for assessing the accuracy of a new theory going forward if only low order corrections are needed.

\section{A Numerical Problem}\label{Sec:Numerical}

In addition to the analytic work above, we have run multi-state calculations on a well-studied model for proton transfer developed by Borgis {\em et al.} \cite{borgis-model-2006} recently studied within a phase-space electronic structure theory by Bian {\em et al.}. \cite{bian_PS-vibration_2025} and Wu {\em et al.} \cite{wu_exact-vibration_2025}. The problem is a three-particle one-dimension model for hydrogen bonding:
\begin{equation}
    \hH = \frac{\hat{P}_1^2}{2M} + \frac{\hat{P}_2^2}{2M} + \frac{\hat{p}_e^2}{2m} + \hat{V}(\hat{r}_e,\hat{R}_1,\hat{R_2})
\end{equation}
In order to avoid the (boring) problem of center of mass motion, we begin by transforming to internal coordinates:
\begin{align}
    \hat{R}_{MCM} = \frac{M\hat{R}_1 + M\hat{R}_2 + m\hat{r}_e}{2M+m}, \quad
    \hat{R}=\hat{R}_1-\hat{R}_2, \quad
    \hat{r} = \hat{r}_e - \frac{\hat{R}_1 + \hat{R}_2}{2}
\end{align}
This transformation allows us to write the Hamiltonian as
\begin{align}
    \hat{H} &= \frac{\hat{P}_{MCM}^2}{2(2M+m)}  + \frac{\hat{P}^2}{2\mu} + \frac{\hat{p}^2}{2m} + \frac{\hat{p}^2}{2(M+M)} + \hat{V}(\hat{R},\hat{r})\\
    &= \frac{\hat{P}_{MCM}^2}{2(2M+m)} + \frac{\hat{P}^2}{2\mu} + \frac{\hat{p}^2}{2m^*} + \hat{V}(\hat{R},\hat{r})
\end{align}
where we have defined the effective mass $m^*= \frac{2Mm}{2M+m}$ arising from including the mass polarization term. As suggested above,  we will neglect (the uncoupled) center of mass motion to obtain an effective 2-particle model. The potential energy term can  be succinctly written in the internal coordinates:
\begin{align}
    \hat{V}(\hat{R},\hat{r}) &= D \left(e^{-2a\left(\frac{\hat{R}}{2} + \hat{r} - d \right)} - 2e^{-a\left( \frac{\hat{R}}{2} + \hat{r} - d \right)} + 1  \right)\\
    &\quad +Dc^2\left( e^{-\frac{2a}{c}\left(\frac{\hat{R}}{2} - \hat{r} - d\right)} 
    -2e^{-\frac{a}{c}\left( \frac{\hat{R}}{2} - \hat{r} - d \right)} \right)\\
    &\quad +Ae^{-B\hat{R}^2} - \frac{C}{\hat{R}^6}
\end{align}
The parameters used are given in Table $\ref{tab:borgis}$. 
\begin{table}[]
    \centering
    \begin{tabular}{|c|c|c|c|c|c|c|}
    \hline
        $D$ & $d$ & $a$ & $c$ & $A$ & $B$ & $C$ \\ \hline
        60 kcal/mol & 0.95 \AA & 2.52 \AA$^{-1}$& 0.707 & $2.32\times10^{5}$ kcal/mol & 3.15 \AA$^{-1}$ & $2.31\times 10^{4}$ kcal/mol/\AA$^6$\\   \hline
    \end{tabular}
    \caption{Model parameters used from Ref. \cite{borgis-model-2006}}
    \label{tab:borgis}
\end{table}

The Hamiltonian above can be directly and exactly diagonalized. To that end, exact calculations were performed by directly constructing the joint nuclear-electronic Hamiltonian with nuclear kinetic energy constructed through Fourier transform of the conjugate momentum grid. We then used a Davidson algorithm to approximately converge the lowest two vibronic states with a convergence threshold of 10$^{-12}$ hartree energy deviation and 10$^{-6}$ hartree for the residual.  
Let us now describe very briefly the nature of the approximate calculations reported below.

\subsection{Born-Huang Calculations}

For the Born-Huang based calculations (labeled as BH), we diagonalize the partially Wignerized Hamiltonian, ignoring the molecular center of mass kinetic energy:
\begin{equation}
    \hH_W(R,P) = \frac{P^2}{2\mu} + \frac{\hat{p}^2}{2m^*} + \hat{V}_W(\hat{r};R) = \hL_W(R)\hLambda_W(R,P)\hL_W^\dagger(R)
\end{equation}
An inverse Weyl transform of the lowest eigensurface yields the Born-Oppenheimer vibrational Hamiltonian:
\begin{equation}
    \hH^{BO} = W^{-1}\left[\left( \hLambda_W\right)_{00}\right]
\end{equation}

For multi-state Born-Huang calculations, one diagonalizes the electronic Hamiltonian  and keeps multiple eigenvalues at each point $(R,P)$ in phase space. Furthermore, one must be sure to impose a consistent phase on the eigenstates so that they change smoothly. While BH(2) theory is formally exact for a complete set of states, BH(1) is not exact and results can depend on the choice of phase.  To that end,  parallel transport is applied before the derivative coupling $\hD_R$ is evaluated. Thereafter, it is straightforward to construct $H^{BH}_W$ via Eq. \ref{eq:MS-BH-main} with either only the first order correction (labeled as BH(1)) or both the first and second order correction (labeled as BH(2)). To avoid spurious inclusion of derivative couplings from the virtual space, $\hD_R$ is first projected into the active space of electronic states before constructing the 2nd order correction. Lastly, for a  system with $N_e$ electronic states expressed on a grid with $N_R$ points, we Weyl transform the $N_R N_r \times N_R N_r$ matrix 
$\hat{H}_W^{BH}$ from an (R,P)-dependent symbol back to an operator with matrix elements, $\left< R\middle| \hat{H}^{BH} \middle| R' \right>$ , which can then be diagonalized.

\subsection{PSEST Calculations}
For phase space electronic structure based calculations, we start with the following PS ansatz:
\begin{equation}
    \hH_W^{PS} 
    = \frac{(P-i\hbar \hGamma)^2}{2\mu} + \frac{\hat{p}^2}{2m^*} + \hat{V}_W(\hat{r};R) 
    = \hL^{PS}_W(R,P) \hLambda^{PS}_W(R,P) \hL_W^{PS\dagger}(R,P)
\end{equation}
For this problem, we use the following $\hGamma$ operator previously used for this problem by Bian {\em et al.}. \cite{bian_PS-vibration_2025} and Wu {\em et al.}. \cite{wu_exact-vibration_2025}:
\begin{gather}
    \hGamma = \frac{\hGamma_1 - \hGamma_2}{2}\\
    \hGamma_1 = \frac{1}{2i\hbar}\left( \hat{\theta}_1 \hat{p} + \hat{p}\hat{\theta}_1 \right), \quad \hGamma_2 = \frac{1}{2i\hbar}\left( \hat{\theta}_2 \hat{p} + \hat{p}\hat{\theta}_2 \right)\\
    \hat{\theta}_1 = \frac{e^{-\big|\hat{r} - \frac{R}{2}\big|^2 /\sigma^2}}{e^{-\big|\hat{r} - \frac{R}{2}\big|^2 /\sigma^2} + e^{-\big|\hat{r} + \frac{R}{2}\big|^2 /\sigma^2}}, \quad \hat{\theta}_2 = \frac{e^{-\big|\hat{r} + \frac{R}{2}\big|^2 /\sigma^2}}{e^{-\big|\hat{r} - \frac{R}{2}\big|^2 /\sigma^2} + e^{-\big|\hat{r} + \frac{R}{2}\big|^2 /\sigma^2}}
\end{gather}

\subsubsection{Single-State Phase Space Results}
To obtain single-state phase-space results, we inverse Weyl transform the lowest phase-space eigensurface to produce an effective vibrational Hamiltonian.
\begin{equation}
    \hH^{SS-PS} = W^{-1}\left[\left(\hLambda_W\right)_{00}\right]
\end{equation}
We then diagonalize $H^{SS-PS}$ to obtain vibrational energies.

\subsubsection{PS-LF calculations}
For PS-LF calculations, we evaluate the corrections summarized in Eq. \ref{eq:PS-LF-master-eq}. Thereafter, just as for a multi-state BH problem, $\hat{H}_W^{PS-LF}$ is Weyl transformed from an (R,P)-dependent symbol back to an operator with matrix elements, $\left< R\middle| \hat{H}^{PS-LF} \middle| R' \right>$ , which can then be diagonalized.

\subsubsection{PS-CF calculations}
To obtain PS-CF corrections, we first rotate the phase-space wavefunctions with the corresponding similarity transform. 
\begin{gather}
    \hat{\tilde{H}}_W (R,P) = \hU_W(R) \hH^{PS}_W(R,P) \hU^\dagger_W(R) =  \hU_W \hL^{PS}_W \hLambda^{PS}_W \hL^{PS\dagger}_W \hU^\dagger_W = \hY_W\hLambda_W\hY_W^\dagger\\
    \hat{U}_W(R) = \mathcal{P} \exp\left( -\int_{R_0}^{R} \hGamma(R') dR'\right)
\end{gather}
Here, $\mathcal{P}$ denotes path ordering. Numerically, we use a midpoint propagator to evaluate $\hU_W(R)$ starting from $R_0=2\text{\AA}$:
\begin{equation}
    \hU_W(R+\Delta R) = \hU_W(R) \exp\left( -\frac{\hGamma(R+\Delta R) + \hGamma(R)}{2}  \Delta R\right), \  \hU_W(R_0)=\hat{I}
\end{equation}

Using $\hU_W(R)$, the rotated wavefunctions $\hY_W$ satisfy:
\begin{equation}
    \hY_W(R,P) = \hU_W(R)\hL^{PS}_W(R)
\end{equation}

We then evaluate the PS-CF corrections summarized in Eq. \ref{eq:PS-CF-master-eq}. From this point on, the calculations proceed through a procedure similar to the Born-Huang or PS-LF calculations described above.

\subsection{Grid Details}

For all calculations, we used a nuclear grid $R \in [2,6] \text{\AA}$ with $N_R=400$ grid points and electronic grid $r\in [-3,3] \text{\AA}$  with $N_e=400$ grid points. All calculations use a nuclear momentum grid $P \in [-\pi/\Delta R, \cdots, \pi(N_P-2)/N_p\Delta R]$ with $N_P=400$ grid points. Electronic kinetic energy operators were constructed with the finite-difference stencil technique. For PS electronic structure calculations, we choose $\sigma=0.5$ bohr$^{-1}$. 

\subsection{Results}\label{Sec:Results}

All of the theory above was implemented and is presented in Fig. \ref{fig:borgis} below. Several conclusions are clear. First, if we include only one electronic state, then just as found previously in Refs. \cite{bian_PS-vibration_2025}, a SS-PS calculation outperforms a BO calculation. Second, if we are to include only two or three electronic states, then the PS-CF(2) approach  gives by far the best absolute ground and excited state energies when compared to the other approaches. Notably, however, 
one can still recover an accurate, relative energetic {\em gap} from a PS calculation according to the lower panel of Fig. \ref{fig:borgis}. 
Third, when one includes enough electronic states, the BH(2) error gets very small as it must; after all, BH(2) is exact in the limit of a complete set of electronic states. That being said, the error in the PS-CF(2) and PS-LF(2) algorithms converges to a small but fixed number as the number of electronic states increases -- because neither PS-CF(2) nor PS-LF(2) is exact in the limit of a complete set of electronic states.  Fourth and lastly, one does not find a big improvement using a first order correction with either a  BH(1) or PS(1) calculation; a second order correction would appear to be needed.

This last point deserves a few more words of context, especially in the context of Ref. \cite{littlejohn-flynn_perturbation-theory_1991} and Ref. \cite{wu_exact-vibration_2025}. In particular, one can note that Ref. \cite{wu_exact-vibration_2025} did find a powerful correction at first order. What's the difference between that result and the present paper? To that end, here it is crucial to emphasize that, in the present paper where we sought a multi-state representation of the electronic problem, we set $A = 0$  in Eq. \ref{eq:L-1st}; by contrast, in Ref. \cite{wu_exact-vibration_2025}, one chose $A$ so as to diagonalize the electronic Hamiltonian. As a result, while we were able here to achieve only minimal results in first order, Ref. \cite{wu_exact-vibration_2025} achieved strong results at first order.

Now, the rationale for our ignoring $A$ in Eq. \ref{eq:L-1st} was our hope that, if we did not diagonalize the Hamiltonian in Eq. \ref{eq:PS-LF-master-eq} before Weyl-transformation, the resulting PS-LF and PS-CF results would be stable near an avoided crossing, where two electronic states come close together in energy.  After all, for the approach in Ref. \cite{wu_exact-vibration_2025}, the energy gap (between the state of interest and all other states) appears at third order for the first-order energy correction. By contrast, here the energy gap appears at second-order only, so that our approach should be more stable than found in Ref. \cite{wu_exact-vibration_2025}.  Moreover, by selecting for $A$, Ref. \cite{wu_exact-vibration_2025} requires a privileged subspace, while here we need not do so. 
Thus, while the present approach will never be as  stable as Born-Huang theory, we have found it to be stable over a reasonably broad range of parameters. Finally and perhaps more importantly, if instability arises, Eqs. \ref{eq:PS-LF-main} (and \ref{eq:PS-CF-main}) can always be cast into a diabatic basis, where convergence should be possible. 
To do so, if we choose  $\hU^D_W(R,P)$ to be an adiabatic-to-diabatic transformation, then we can follow the approach of  Ref. \cite{Zaidi-electron_transfer-2026} and work in the relevant diabatic framework: 
\begin{align}
    \hH^{PS}_W &=  \hL_W^{PS} \hLambda_W^{PS} \hL^{PS\dagger}_W \\
    &= \left(\hL_W^{PS} \hU^D_W \right) \left( \hU_W^{D\dagger}\hLambda_W^{PS} \hU^D_W \right) \left( \hU_W^{D\dagger} \hL^{PS\dagger}_W \right)\\
    &= \hL^{D,PS}_W \hLambda^{D,PS}_W \hL_W^{D,PS\dagger}
\end{align}
$\hLambda^{D,PS}_W$ are the diabatic energies and $\hL^{D,PS}_W$ are the corresponding diabatic wavefunctions.

\begin{figure}
    \centering    \includegraphics[width=1.0\linewidth]{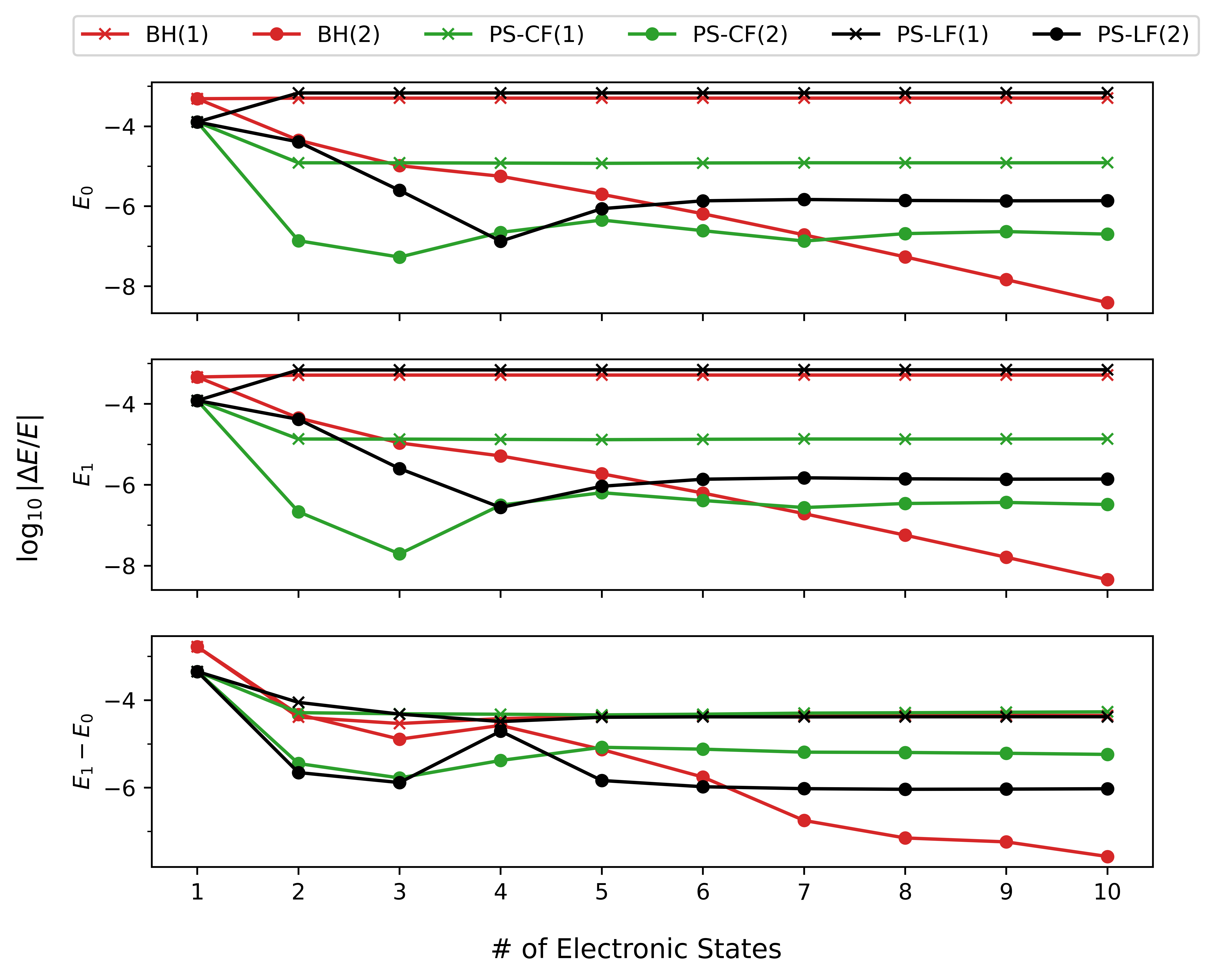}
    \caption{Relative error on a log scale for the ground state vibronic energy (top), first excited state vibronic energy (middle), and lowest vibronic energy gap (bottom) for the Borgis hydrogen bonding model \cite{borgis-model-2006} as a function of number of electronic states included.
    Here we plot the 1st order Born-Huang [BH(1)] and the 2nd order Born-Huang [BH(2)] results.  Because the latter is exact, one necessarily finds that the results approach the exact Hamiltonian as one increases the number of electronic states. 
    (The BH(1) or BH(2) are exactly equal to the BO energy when the number of electronic states is 1.) 
    Here we also plot 1st and 2nd order PS-LF results (PS-LF(1) and PS-LF(2)) as well as first and second order PS-CF  (PS-CF(1) and PS-CF(2)) results.  Unlike BH theory, the PS calculations are not exact at any finite order; the exact result can be obtained only by going to the infinite order and reporting  PS-LF($\infty$) or  PS-CF$(\infty$) results. That being said, for  a small number of electronic states, we find that PS-LF(2) and PS-CF(2) routinely outperform all BH calculations. All calculations are performed at a mass ratio of $M/m=200$.} 
    \label{fig:borgis}
    
\end{figure}

\section{Discussion: One  Vs. Three
Dimensions}\label{Sec:Discussion}

Above, we have presented a reasonably complete theory of how to express the total nuclear-electronic Hamiltonian for a given $\hbGamma$ operator. Perhaps most interestingly, we have shown that a very powerful expression can be utilized when $\hbGamma$ has zero non-Abelian curl and can be written as $-\hU_W^\dagger \partial_R \hU_W$ so that we can use the PS-CF strategy above.
This finding begs the question: when and if and how can such a  $\hbGamma$ operator be constructed?
While a curl-free condition can be imposed in one dimension, it would appear impossible in two or three dimensions most generally.

\subsection{One Dimension}

Let us first show that our previously proposed\cite{tao_basis-free_2025} choice of $\hGamma$ in 1D systems  yields zero non-Abelian curvature. In this case, our choice of $\hGamma$ operator is a symmetrized product of the electronic momentum operator and a partition of unity function $\hat{\theta}_A$:
\begin{equation}
    \hGamma_A = \frac{1}{2i\hbar}\acom{\hat{\theta}_A}{\hat{p}},\quad \sum_A \hat{\theta}_A = \hat{1}
\end{equation} 
Moreover, for a chain of atoms in one dimension, $\hat{\Omega}$ depends only on the atom index:
\begin{equation}\label{eq:1D-nonabelian}
    \hat{\Omega}^{A,B} = \frac{\partial \hat{\Gamma}_{B}}{\partial R_{A}} - \frac{\partial \hat{\Gamma}_{A}}{\partial R_{B}} -  \com{\hat{\Gamma}_{A}}{\hat{\Gamma}_{B}}
\end{equation}

To show that Eq. \ref{eq:1D-nonabelian} vanishes for all atoms and a given choice of $\hat{\theta}$, note that the commutator term can be written as:
\begin{align}
    \com{\hGamma_A}{\hGamma_B}
    &= -\frac{1}{2i\hbar}\acom{\hat{J}_{AB}}{\hat{p}}\\
    \hat{J}_{AB} &=  \hat{\theta}_A\hat{\theta}_B' - \hat{\theta}_B \hat{\theta}_A'
\end{align}
where we  define $\hat{\theta}_A'  =(\partial_r \hat{\theta}_A)$. The nuclear derivatives can also be rearranged in an anti-commutator form: 
\begin{align}
    \frac{\partial \hat{\Gamma}_{B}}{\partial R_{A}} - \frac{\partial \hat{\Gamma}_{A}}{\partial R_{B}} &= \frac{1}{2i\hbar}\acom{\hat{\Delta}_{AB}}{\hat{p}}
    \\
    \hat{\Delta}_{AB}& = \partial_{R_A}\hat{\theta}_B - \partial_{R_B} \hat{\theta}_A
\end{align}

Thus, for our choice of $\hGamma$, we find:
\begin{align}
    \hat{\Omega}^{A,B} &= \frac{1}{2i\hbar} \acom{\hat{K}_{AB}}{\hat{p}} = 0\\
    \hat{K} _{AB}&= \partial_{R_A}\hat{\theta}_B - \partial_{R_B} \hat{\theta}_A  + \hat{\theta}_A  \hat{\theta}'_B - \hat{\theta}_B \hat{\theta}'_A
\end{align}

Let us now show that $\hat{K}_{AB}$ vanishes provided that we choose $\hat{\theta}$ to be a function of the electronic coordinate relative to the nuclear coordinates, 
\begin{equation}
    \hat{\theta}_A = \hat{\theta}(\hat{r}-R_A,\hat{r}-R_B,\cdots) 
    \label{eq:relative}
\end{equation}
and we insist that only two $\theta$ functions can be non-zero at any point in space. In such a scenario, 
    \begin{align}
        \hat{\theta}_A + \hat{\theta}_B = 1 \implies \hat{\theta}_B=1-\hat{\theta}_A
        \label{eq:twonz}
    \end{align}
A simple computation using Eq. \ref{eq:twonz} then shows that
\begin{align}
    \hat{K} _{AB}&= \partial_{X_A}\hat{\theta}_B - \partial_{X_B} \hat{\theta}_A  + \hat{\theta}_A  \hat{\theta}'_B - \hat{\theta}_B \hat{\theta}'_A \\
    &= \partial_{X_A}\hat{\theta}_B - \partial_{X_B} \hat{\theta}_A  - (1- \hat{\theta}_B)  \hat{\theta}'_A - \hat{\theta}_B \hat{\theta}'_A \\
    &= \partial_{X_A}\hat{\theta}_B - \partial_{X_B} \hat{\theta}_A  - \hat{\theta}'_A \\
    &= - \partial_{X_A}\hat{\theta}_A - \partial_{X_B} \hat{\theta}_A  - \hat{\theta}'_A
    \\
    &= 0
\end{align}
where the very last line follows from Eq. \ref{eq:relative}.

\subsection{Three Dimensions}
The proof above for a curl-free system is very one-dimensional, and all of our existing expressions for $\hbGamma$ that satisfy the relevant translational and rotational constraints   for two and three dimensional problems (see Eqs. 99-101 in Ref. \cite{bhati-MSW-2026}) exhibit non-vanishing curls. For example, as shown by Moody, Shapere, and Wilczek \cite{MSW-1986},  the non-Abelian curvature for a diatomic molecule in 3D is nonzero due to the non-commutativity of rotations.  Moreover, as shown by Bhati {\em et al.}. \cite{bhati-MSW-2026}, our choice of $\hat{\Gamma}$ mimics the exact non-Abelian curvature and is also nonzero for a diatomic molecule in 3D system, so that constructing a curl-free $\hbGamma$ may not even be desirable. That being said, one must wonder if it is possible to construct a meaningful $\hbGamma$ operator in 3D that offers improved energies while also maintaining a zero non-Abelian curl. This question will need to be addressed as soon as possible, and help from mathematicians may be necessary. 

\section{Conclusions}\label{Sec:Conclusion}
We have presented here a thorough investigation of nuclear-electronic dynamics within the lens of  a non-Born-Huang phase space electronic structure framework. Using Wigner-Weyl transformations and star products, and basing our analysis on previous work by Littlejohn and Flynn, we have shown exactly how to represent the total nuclear+electronic Hamiltonian within a basis of electronic states that are parametrized by both $R$ and $P$. Our analysis extends previous work by Wu {\em et al.} by demonstrating exactly how to construct the relevant Hamiltonian when we include more than one electronic state  in the final representation.
We have further analyzed two separate cases: the case where $\hbGamma$ is curl free and can be written as $-\hU_W^\dagger \partial_R \hU_W$ (which allows for further unitary transformations and simplifications), and the case where $\hbGamma$ is not curl free and no further simplification is possible.  For a one-dimensional problem, both approaches above are valid, though preliminary evidence suggests that, when possible, using the curl-free condition can lead to more robust results. PS results always outperform BH results when we use only a few electronic states.

Of all the results presented here, perhaps the most interesting finding is the fact that, when $\hGamma = -\hU_W^\dagger \partial_R \hU_W$, one can write down Eq. \ref{eq:PS-CF-master-eq}, which makes clear that using a phase space electronic representation is entirely equivalent to using a BH framework -- with the only change being that one invokes a preconditioner $\hU$ that should hopefully make convergence faster. Given this finding, the most outstanding question is whether or not such $\hU$ can be found for multidimensional problems. In Sec. \ref{Sec:Discussion} above, we have shown that $\Gamma$ operators can always be constructed in a curl-free fashion in 1D, but we are not aware of any such constructions that are meaningful in 3D.

Altogether, the theory and results presented here highlight both (i) the power of novel  non-Born Oppenheimer electronic structure approaches to deliver new information about coupled nuclear-electronic systems as well   as (ii) the need for a deeper understanding of how best to build such a PS approach (especially given the fact that theoretical analysis above in Sec. \ref{Sec:PS-LF} has not relied on any specific form for the $\hbGamma$ operator). Of course, there is also now the opportunity to extend the present approach to a new and powerful phase-space approach\cite{bian:2024:pssh_translations_rotations} to nonadiabatic dynamics, which can be compared against more traditional semiclassical approaches (using either surface hopping or Ehrenfest like methods)\cite{tully:fssh,coker:2008:iterative} or methods based on the quantum-classical Liouville equation\cite{martens:1997:partwig,kapral:1999:jcp,coker:2008:iterative,coker:2012:iterative}.  Our hope is that the present manuscript will prove helpful towards the latter goals.

\section{Data Availability}
The data that support the findings of this study are available from the corresponding author upon reasonable request. 

\section{Acknowledgments}
The authors would like to thank Xinchun Wu, Nadine Bradbury, and Tim Duong for their helpful discussion. We also thank Prof. Steven Wiggins for suggesting that we include the material in Appendix \ref{app:response} in this draft. 
This work was supported by the U.S. Air Force Office of Scientific Research (USAFOSR) under Grant No. FA955023-1-0368. Y.F. was supported by the Fulbright Program.

\appendix

\section{Derivation of 2nd order Born-Huang Term}\label{app:2ndOrderBH}
Here, we show how to extract the relevant 2nd order Born-Huang terms in Eq. \ref{eq:newBH} starting from Eq. \ref{eq:lambda-2nd}. Note that because Born-Huang wavefunctions are only a function of nuclear position $R$, wavefunction corrections are zero, and therefore only the third line of Eq. \ref{eq:lambda-2nd} contributes to the 2nd order Born-Huang correction. 
For convenience, let us drop the subscript $W$.  If we keep only 2nd order derivatives with respect to $\hLambda$ when expanding the 2nd order derivatives of $\hY \hLambda \hY^\dagger$ in Eq. \ref{eq:lambda-2nd}, the result is:
\begin{equation}
    \begin{aligned}
        \hY^\dagger\dpb{\hY \hLambda \hY^\dagger }{\hY} &= \hY^\dagger\left( (\hY\hLambda \hY^\dagger)_{RR}\hY_{PP} + (\hY\hLambda \hY^\dagger)_{PP}\hY_{RR} - 2(\hY\hLambda \hY^\dagger)_{RP}\hY_{RP} \right)\\
        &\approx \hY^\dagger\left( \hY \hLambda_{RR} \hY^\dagger \hY_{PP} + \hY \hLambda_{PP}\hY^\dagger \hY_{RR} - 2\hY\hLambda_{RP}\hY^\dagger \hY_{RP}\right)\\
        &= \hLambda_{RR} \hY^\dagger \hY_{PP} + \hLambda_{PP} \hY^\dagger \hY_{RR} - 2\hLambda_{RP}\hY^\dagger \hY_{RP}
    \end{aligned}
\end{equation}

\begin{equation}
    \begin{aligned}
        \dpb{\hY^\dagger}{\hY\hLambda} &= \hY^\dagger_{RR}(\hY\hLambda)_{PP} + \hY^\dagger_{PP}(\hY\hLambda)_{RR} - 2\hY^\dagger_{RP}(\hY\hLambda)_{RP}\\
        &\approx \hY^\dagger_{RR}\hY\hLambda_{PP} + \hY^\dagger_{PP}\hY\hLambda_{RR} - 2\hY^\dagger_{RP}\hY\hLambda_{RP}
    \end{aligned}
\end{equation}

\begin{equation}
    \begin{aligned}
        \pb{\hY^\dagger}{\pb{\hY\hLambda \hY^\dagger}{\hY}} &= \pb{\hY^\dagger}{(\hY\hLambda \hY^\dagger)_R\hY_P - (\hY\Lambda \hY^\dagger)_P\hY_R}\\
        &\approx \pb{\hY^\dagger}{\hY\hLambda_R \hY^\dagger \hY_P - \hY\hLambda_P\hY^\dagger \hY_R}\\
        &\approx \hY^\dagger_R \hY \hLambda_{RP}\hY^\dagger \hY_P  - \hY^\dagger_R\hY\hLambda_{PP}\hY^\dagger \hY_R\\
        &\quad -  \hY^\dagger_P \hY\hLambda_{RR}\hY^\dagger \hY_P + \hY^\dagger_P \hY\hLambda_{RP}\hY^\dagger \hY_R\\
        &= \hD^\dagger_R\hLambda_{RP}\hD_P -\hD^\dagger_R\hLambda_{PP}\hD_R - \hD^\dagger_P\hLambda_{RR}\hD_P + \hD^\dagger_P\hLambda_{RP}\hD_R
    \end{aligned}
\end{equation}
The sum of these terms gives us the relevant BH 2nd order coupling:
\begin{equation}
    \begin{aligned}
        \hLambda_{\mathrm{2, \ line\ 3}}&\approx \frac{1}{4}\left( -\hD_R\hLambda_{PP}\hD_R -\hD_P\hLambda_{RR}\hD_P + \hD_R\hLambda_{RP}\hD_P + \hD_P\hLambda_{RP}\hD_R\right)\\
        &\quad-\frac{1}{8} \left( \hLambda_{RR}\hY^\dagger \hY_{PP} + \hY^\dagger_{PP}\hY\hLambda_{RR}+ \hLambda_{PP}\hY^\dagger \hY_{RR} + \hY^\dagger_{RR}\hY\hLambda_{PP} -2\left(\hLambda_{RP}\hY^\dagger \hY_{RP} +\hY^\dagger_{RP}\hY\hLambda_{RP}\right) \right)
    \end{aligned}
\end{equation}
We can rewrite this expression in a cleaner form by noting that $\hY^\dagger \hY_{RR} = \partial_R \hD_R + \hD_R^2$,  $\hY^\dagger \hY_{RP} = \partial_R \hD_P + \hD_R\hD_P$, and $\hY^\dagger_{RP}\hY = \hD_P\hD_R - \partial_R\hD_P$:
\begin{equation}
    \begin{aligned}
        \hLambda_{\mathrm{2\ line\ 3}} \approx& \frac{1}{4}\left( -\hD_R\hLambda_{PP}\hD_R -\hD_P\hLambda_{RR}\hD_P +\hD_R\hLambda_{RP}\hD_P + \hD_P\hLambda_{RP}\hD_R\right)\\
        -&\frac{1}{8}\left( \hLambda_{RR} \hD_P^2 + \hD_P^2\hLambda_{RR} + \hLambda_{RR} \partial_P \hD_P - \partial_P \hD_P \hLambda_{RR}\right. \\
        &\quad\quad \left. + \hLambda_{PP}\hD_R^2 + \hD_R^2 \hLambda_{PP} + \hLambda_{PP}\partial_R\hD_R - \partial_R\hD_R\hLambda_{PP} \right. \\
        &\quad\quad \left. - 2 \left( \hLambda_{RP}\hD_R\hD_P + \hD_P\hD_R\hLambda_{RP}  + (\hLambda_{RP} \partial_R\hD_P - \partial_R\hD_P\hLambda_{RP}) \right) \right)\\
    \end{aligned}
\end{equation}
If we rewrite again using commutators $[\cdot,\cdot]_-$ and anticommutators $[\cdot,\cdot]_+$, we find:
\begin{equation}
    \begin{aligned}
    \label{eq:2ndCorrLine3}
        \hLambda_{\mathrm{2\ line\ 3}} \approx& \frac{1}{4} \left( -\hD_R\hLambda_{PP}\hD_R -\hD_P\hLambda_{RR}\hD_P +\hD_R\hLambda_{RP}\hD_P + \hD_P\Lambda_{RP}\hD_R\right)\\
        &\quad -\frac{1}{8}\left( [\hLambda_{PP},\hD_R^2]_+ +[\hLambda_{RR},\hD_P^2]_+ + [\hLambda_{PP},\partial_R \hD_R]_- + [\hLambda_{RR},\partial_P \hD_P]_- \right)\\
        &\quad +\frac{1}{4} \left(\hLambda_{RP} \hD_R\hD_P + \hD_P\hD_R \hLambda_{RP} + [\hLambda_{RP},\partial_R\hD_P]_- \right)
    \end{aligned}
\end{equation}
Now, consider the Born-Huang Hamiltonian $\hH_W=P^2/2M + \hH_{el}(R)$. In this case, $\hD_P=0$, $\partial_PD_R=0$, and $\hLambda_{PP} = 1/M$. All commutators of $\hLambda_{PP}$ vanish.   Thus, Eq. \ref{eq:2ndCorrLine3} does indeed reduce to
\begin{equation}
    \hLambda^{BH}_{2} = -\frac{1}{4}\hD_R\hLambda_{PP}\hD_R -\frac{1}{8}[\hLambda_{PP},\hD_R^2]_+ = -\frac{1}{2M}\hD^2_R,
\end{equation}
which is the expected 2nd order BH correction. 

\section{1st and 2nd order corrections in the PS-LF expansion}\label{app:PS-LF-Lambda}
Sec. \ref{Sec:PS-CF} above treats the PS-CF approach to second order. Here, in this section, we will now show the additional terms that arise when utilizing the PS-LF expansion (as compared to the PS-CF expansion in Eq. \ref{eq:PS-CF-master-eq}). Starting from Eq. \ref{eq:PS-LF-main} to 1st order, the only additional term for PS-LF (relative to PS-CF) is proportional to $\hGamma$:
\begin{equation}
    \Delta \hLambda_1 = i \frac{P}{M} \hL_W^\dagger \hGamma \hL_W = i\frac{P}{M} \hat{\Gamma}_L
\end{equation}
We denote the $\hGamma$ operator in the adiabatic electronic basis as $\hat{\Gamma}_L$. 
The total first order correction is therefore:
\begin{equation}
    \begin{aligned}
        \hLambda_1^{PS-LF} =& i\frac{P}{M}\hat{\Gamma}_L -\frac{i}{2}\left( \hD_R \hLambda_P + \hLambda_P \hD_R -  \hD_P \hLambda_R - \hLambda_R \hD_P \right)  
        \\ & + \frac{i}{2}\left( \hD_R \hLambda \hD_P - \hD_P \hLambda \hD_R \right)
        \\ & - \frac{i}{4}\left( [\hD_R, \hD_P]\hLambda + \hLambda [\hD_R, \hD_P]\right)
    \end{aligned}
\end{equation}
Similarly, the additional terms at 2nd order are:
\begin{equation}
    \begin{aligned}
        \Delta \hLambda_2 =&  \frac{1}{2M} \hL_W^\dagger \hGamma^2 \hL_W + i \frac{P}{M} \hL_1^{\dagger} \hGamma \hL_W + i\frac{P}{M} \hL_W^\dagger \hGamma \hL_1 - \frac{1}{2M} \hL_W^\dagger \pb{P \hGamma}{\hL_W} - \frac{1}{2M} \pb{\hL_W^\dagger}{P\hGamma \hL_W}\\
        =& \frac{1}{2M} \left( \hat{\Gamma}_L^2 + \acom{\hat{\Gamma}_L}{\hD_R} \right) - \frac{P}{4M} \acom{\hat{\Gamma}_L}{\com{\hD_R}{\hD_P}} -\frac{P}{2M}\left( \acom{D_P}{\hL_W^\dagger \hGamma_R \hL_W} + \hD_P\hat{\Gamma}_L\hD_R - \hD_R\hat{\Gamma}_L\hD_P \right)
    \end{aligned}
\end{equation}
We define $\hGamma_R = \frac{\partial \hGamma}{\partial R}$ for brevity.
The total 2nd order correction is then:
\begin{equation}
    \begin{aligned}
        \hLambda_2^{PS-LF} 
        &= \frac{1}{2M} \left( \hat{\Gamma}_L^2 + \acom{\hat{\Gamma}_L}{D_R} \right) - \frac{P}{4M} \acom{\hat{\Gamma}_L}{\com{\hD_R}{\hD_P}} -\frac{P}{2M}\left( \acom{D_P}{\hL_W^\dagger \hGamma_R \hL_W} + D_P\hat{\Gamma}_LD_R - D_R\hat{\Gamma}_LD_P \right)\\
        &+\hL_2^\dagger \hL_W \hLambda_W + \hL^\dagger_1 \hL_W \hLambda_W \hL_W^\dagger \hL_1 + \hLambda_W  \hL_W^\dagger \hL_2 \\
        &+ \frac{i}{2} \left(\hL_W^\dagger\pb{\hL_W \hLambda_W \hL_W^\dagger}{\hL_1} + \hL^\dagger_1\pb{\hL_W \hLambda_W \hL_W^\dagger}{\hL_W} + \pb{\hL^\dagger_1}{\hL_W \hLambda_W } + \pb{\hL_W^\dagger}{\hL_W \hLambda_W \hL_W^\dagger \hL_1}\right)\\
        &-\frac{1}{8} \left( \hL_W^\dagger \dpb{\hL_W \hLambda_W \hL_W^\dagger}{\hL_W} + 2 \pb{\hL_W^\dagger}{\pb{\hL_W \hLambda_W \hL_W^\dagger}{\hL_W}} + \dpb{\hL_W^\dagger}{\hL_W \hLambda_W } \right)
    \end{aligned}
\end{equation}
In summary, the total (corrected) $\hLambda'$ to second order is (including zeroth, first and second order terms):
\begin{equation}\label{eq:PS-LF-master-eq}
    \begin{aligned}
        \hH_{W}^{PS-LF(2)} =& \hLambda_{W}  + i\hbar\frac{P}{M}\hat{\Gamma}_L -\frac{i\hbar}{2}\left( \hD_R \hLambda_P + \hLambda_P \hD_R -  \hD_P \hLambda_R - \hLambda_R \hD_P \right)  + \frac{i\hbar}{2}\left( \hD_R \hLambda_{W} \hD_P - \hD_P \hLambda_{W} \hD_R \right)
        \\ & - \frac{i\hbar}{4}\left( [\hD_R, \hD_P]\hLambda_{W} + \hLambda_{W} [\hD_R, \hD_P]\right)\\
        & + \frac{\hbar^2}{2M} \left( \hat{\Gamma}_L^2 + \acom{\hat{\Gamma}_L}{\hD_R} \right) - \hbar^2\frac{P}{4M} \acom{\hat{\Gamma}_L}{\com{\hD_R}{\hD_P}} \\
        & - \hbar^2\frac{P}{2M}\left( \acom{D_P}{\hL_W^\dagger \hGamma_R \hL_W} + D_P\hat{\Gamma}_LD_R - D_R\hat{\Gamma}_LD_P \right)\\
        &+\hbar^2 \left(\hL_2^\dagger \hL_W \hLambda_W + \hL^\dagger_1 \hL_W \hLambda_W \hL_W^\dagger \hL_1 + \hLambda_W  \hL_W^\dagger \hL_2\right) \\
        &+ \frac{i\hbar^2}{2} \left(\hL_W^\dagger\pb{\hL_W \hLambda_W \hL_W^\dagger}{\hL_1} + \hL^\dagger_1\pb{\hL_W \hLambda_W \hL_W^\dagger}{\hL_W} + \pb{\hL^\dagger_1}{\hL_W \hLambda_W } + \pb{\hL_W^\dagger}{\hL_W \hLambda_W \hL_W^\dagger \hL_1}\right)\\
        &-\frac{\hbar^2}{8} \left( \hL_W^\dagger \dpb{\hL_W \hLambda_W \hL_W^\dagger}{\hL_W} + 2 \pb{\hL_W^\dagger}{\pb{\hL_W \hLambda_W \hL_W^\dagger}{\hL_W}} + \dpb{\hL_W^\dagger}{\hL_W \hLambda_W } \right)
    \end{aligned}
\end{equation}

\section{The nuclear-nuclear susceptibility for the different approaches}\label{app:response}

In Sec. \ref{Sec:Analytical}  above, we have analyzed the performance of a phase space electronic structure approach  to recover the lowest vibronic gap. In this section, we will now extend that analysis to a different dynamical variable, namely the nuclear-nuclear frequency dependent susceptibility. With regards to multi-state expansions, we will include only  the first order multi-state correction. Additionally, due to the complexity of the PS-LF(1) expression, we will limit our analysis to a comparison between BH(1) and PS-CF(1). 

The Kubo Formula for the  nuclear-nuclear response function is \cite{kubo1991statistical2}:
\begin{equation}\label{eq:Kubo}
    \chi_{RR}(t) = -\frac{i}{\hbar} \Theta(t) \left< \com{\hat{R}(t)}{\hat{R}(0)} \right>
\end{equation}
Here, $\Theta(t)$ is the Heaviside step function enforcing causality. Now, note that (for $t>0$), the response function has the same equation of motion as does $\hat{R}(t)$. Thus, if we can evaluate $\hat{R}{(t)}$, we can also  evaluate  $\chi_{RR}(t)$. Moreover,   Fourier transforming $\chi_{RR}(t)$ then yields the frequency dependent susceptibility $\tilde{\chi}_{RR}(\nu)$. Finally, because the total Hamiltonian (Eq. \ref{eq:anal-H}) is quadratic, it is simple to 
evaluate $\hat{R}(t)$; after all, the classical and quantum propagation equations are identical for harmonic oscillators. In other words, to construct the frequency dependent susceptibility $\tilde{\chi}_{RR}(\nu)$, we really need only guess $R(t) = R_0e^{i\nu t}$ and  $r(t) = r_0e^{i\nu t}$, drive $R(t)$ by a force $f_0 e^{i\nu t}$, and evaluate $\tilde{\chi}_{RR}(\nu) = R_0/f_0$.

To that end, let us write down the relevant Hamiltonians from above:
\begin{align}
    \label{eq:resp-ex-H}
    \hH^{exact} &= \frac{\hat{P}^2}{2M} + \frac{\hat{p}^2}{2m} + \frac{K}{2} \hat{R}^2 + \frac{g}{2}(\hat{r} - \hat{R})^2
    \\
    \hH^{BO} &= \frac{\hat{P}^2}{2M}+  \frac{1}{2}K\hat{R}^2 +\hbar \sqrt{\frac{g}{m}} \left( \hat{n} + \frac{1}{2}\right)\\
    &= \frac{\hat{P}^2}{2M} + \frac{1}{2}K \hat{R}^2 + \frac{\hat{p}^2}{2m} + \frac{1}{2} g \hat{r}^2 \\
    \hH^{SS-PS} &= \frac{\hat{P}^2}{2(M+m)}+  \frac{1}{2}K\hat{R}^2 + \hbar \sqrt{\frac{g}{\mu}}\left(\hat{n} + \frac{1}{2}\right)
    \\ 
    &= \frac{\hat{P}^2}{2(M+m)}+  \frac{1}{2}K\hat{R}^2 + \frac{\hat{p}^2}{2\mu} + \frac{1}{2} g\hat{r}^2
    \\
    \hH^{BH(1)} &= \frac{\hat{P}^2}{2M}+  \frac{1}{2}K\hat{R}^2 + \frac{\hat{p}^2}{2m} + \frac{1}{2} g\hat{r}^2 - \frac{\hat{P}}{M}\hat{p}
    \\
    \hH^{PS-CF(1)} &= \frac{\hat{P}^2}{2(M+m)}+  \frac{1}{2}K\hat{R}^2 + \frac{\hat{p}^2}{2\mu} + \frac{1}{2} g\hat{r}^2 - \frac{\mu}{M}K \hat{R}\hat{r}
\end{align}

The relevant susceptibilities are then exactly:

\begin{align}
    \label{eq:exact-resp}
    \tilde{\chi}^{exact}_{RR}(\nu) &= \left(K - \nu^2\left(M+ \frac{m}{1-\frac{m}{g}\nu^2}\right)\right)^{-1}
    \\
    \tilde{\chi}^{BO}_{RR}(\nu) &= \left( K-\nu^2M \right)^{-1}
    \\
    \tilde{\chi}^{SS-PS}_{RR}(\nu) &= \left(K - (M+m)\nu^2\right)^{-1}
    \\
    \tilde{\chi}_{RR}^{BH(1)}(\nu) &= \frac{g\left(1 - \frac{m}{M}\right) - m\nu^2}{\left(K-M\nu^2\right)\left(g-m\nu^2\right) - \frac{m}{M}gK}\\
    \tilde{\chi}_{RR}^{PS-CF(1)}(\nu) &= \left(K-\left(M+m\right)\nu^{2}-\frac{\left(\frac{\mu}{M}K\right)^{2}}{g-\mu  \nu^{2}}\right)^{-1}
\end{align}

\subsubsection{Low frequency limit ($\nu \rightarrow 0$)}
To better understand the results above, let us take the low frequency, nearly static limit.
If we expand the exact, PS-CF(1), and BH(1) susceptibilities as a function of $\nu$, we find:

\begin{align}
    \label{eq:susc-ex-taylor}
    \tilde{\chi}_{RR}^{exact}(\nu) &=  \left(K - (M+m)\nu^2 - \frac{m^2}{g} \nu^4 + O\left(\nu^6\right)\right)^{-1}
    \\
    \tilde{\chi}^{BO}_{RR}(\nu) &= \left( K-\nu^2M \right)^{-1}
    \\
    \tilde{\chi}^{SS-PS}_{RR}(\nu) &= \left(K - (M+m)\nu^2\right)^{-1}
    \\
    \tilde{\chi}_{RR}^{BH(1)} (\nu) &= \left(K - \left(M+m+\frac{m^2}{M-m}\right)\nu^2 - \frac{1}{g}\left(\frac{Mm}{M-m}\right)^2 \nu^4 + O\left(\nu^6\right)\right)^{-1}
    \\
    \tilde{\chi}_{RR}^{PS-CF(1)}(\nu) &= \left(\left(K-\frac{K^2\mu^2}{gM^2}\right)-\left(M+m+\frac{K^2\mu^3}{g^2M^2}\right)\nu^2-\left(\frac{K^2\mu^4}{g^3M^2}\right)\nu^4 + O\left(\nu^6\right)\right)^{-1}
\end{align}

Several interesting points arise from the above formulae.  The BO susceptibility fails to reproduce the combined second order $M+m$ mass. However, SS-PS susceptibility does reproduce the correct second order term near zero frequency and the exact response when $g/K \rightarrow \infty$. BH(1) retains the correct static susceptibility, but erroneously includes a $m^2/(M-m)$ term in the second order term whose (erroneous) contribution scales as $O(\eta)$.  The PS-CF(1) susceptibility fictitiously renormalizes the static susceptibility and second order term. However, the error at zero frequency scales as $\sim O(\eta^2)$ and the error of the second order term scales as $\sim O(\eta^3)$, which likely explains why it performs so well in practice. None of the above methods 
recovers the correct fourth order term.

\subsubsection{The upper pole $\omega_+$}
From the data above, one might actually  guess that the single surface phase space approach is actually the best approximation: it is exact at zeroth and second order in $\nu$ and no method is accurate at $\nu^4$. That being said, that is clearly not true, a fact that the can be ascertained by focusing on the upper pole $\omega_+$. Note that, because the Hamiltonian in Eq. \ref{eq:resp-ex-H} can always be decoupled into two independent oscillators, it is clear that the exact response function should yield two poles, the lower and upper poles in Eq. \ref{eq:exact-resp} above.
We have already discussed the accuracy of the lower pole in Sec. \ref{Sec:Analytical}. Let us now  focus on the upper pole. Because the BO and SS-PS susceptibilities do not couple the electronic and nuclear oscillators, they actually lack an upper pole. However, for BH(1) and PS-CF(1), expanding the upper pole with respect to $\eta=m/M$ yields:

\begin{align}
    \omega_{+}^{exact} &= \sqrt{\frac{g}{m}}\left[ 1 + \frac{\eta}{2} + \left(\frac{K}{2g} - \frac{1}{8}\right) \eta^2 + \cdots \right]\\
    \omega_{+}^{BH(1)}  &= \sqrt{\frac{g}{m}} \left[ 1 + \frac{K}{2g}\eta^2 + \cdots \right]\\
    \omega_{+}^{PS-CF(1)} &= \sqrt{\frac{g}{m}} \left[ 1 + \frac{\eta}{2} - \frac{1}{8} \eta^2 + \cdots \right]
\end{align}

Here, we observe a similar separation of spring constant dependent (the ``dynamic") and independent (``static") terms. The BH(1) susceptibility correctly obtains the ``dynamic" second order term, but lacks the first and second order ``static" terms, leading to an error scaling at order $O(\eta)$. Unlike the lower pole, the upper pole retains only the static first and second order terms. Since the first order term is purely static, the PS-CF(1) upper pole error instead scales as order $O(\eta^2)$, which ultimately explains by PS-CF(1) performed well above.

\section{Expansion of the 2nd order energy correction for PS-CF}\label{App:2ndOrderExpansion}
Finally, because the relevant equations at second order for PS-CF are messy and can be difficult to implement, we will 
expand out Eq. \ref{eq:lambda-2nd} in terms of derivative couplings $\hD_R,\hD_P$ and derivatives of the derivative couplings (which is indeed how we implemented the relevant code). 

For notational simplicity, we define the following:
\begin{gather}
    \hD_{RR} = \frac{\partial \hD_R}{\partial R}, \quad
    \hD_{PR} = \frac{\partial \hD_P}{\partial R}, \quad
    \hD_{PP} = \frac{\partial \hD_P}{\partial P}\\
    \hat{C} = \com{\hD_R}{\hD_P}\\ 
    \hat{C}_R = \frac{\partial \hat{C}}{\partial R}, \quad
    \hat{C}_P = \frac{\partial \hat{C}}{\partial P}
\end{gather}

To begin with, note that we can write the 2nd order wavefunction correction (
Eq. \ref{eq:Y-2nd}
) in terms of the above derivatives.

\begin{align}
    \hY_2 &= \hY_W\left( -\frac{3}{32} \hat{C}^2 + \frac{1}{16}\left( \acom{\hat{C}_R}{\hD_P} - \acom{\hat{C}_P}{\hD_R} \right) \right. \nonumber \\
    &\quad \left. + \frac{1}{16} \left(\acom{\hD_R^2}{\hD_P^2} + \com{\hD_R^2}{\hD_{PP}} + \com{\hD_P^2}{\hD_{RR}} - \acom{\hD_{RR}}{\hD_{PP}}\right) \right. \\
    &\quad  \left. -\frac{1}{8} \left( \hD_P \hD_R^2 \hD_P + \hD_P \hD_R\hD_{PR} - \hD_{PR} \hD_R \hD_P - \hD_{PR}^2 \right) \right) \nonumber\\
    &\equiv \hY_W \hat{K}\nonumber
\end{align}
We define $\hat{K}$ in the last line as a shorthand for brevity.

Let us now express Eq. \ref{eq:lambda-2nd} line-by-line. The first line can be simply written as: 
\begin{align}
    \hLambda_{2\text{, line 1}} &= \hat{K}^\dagger \hLambda_W + \hLambda_W \hat{K} - \frac{1}{16} \hat{C} \hLambda_W \hat{C} \\
    &= \acom{\hat{K}}{\hLambda_W} - \frac{1}{16} \hat{C}\hLambda_W \hat{C}
\end{align}

For the second line, if $\hat{X}$ is an operator expressed as a matrix in the phase space adiabatic basis, it will be  helpful to define covariant derivatives (where $I=R,P$):
\begin{gather}
    \nabla_I (\hat{X}) = \frac{\partial \hat{X}}{\partial I} + \com{\hD_I}{\hat{X}}\\
    L_I(\hat{X}) = \frac{\partial \hat{X} }{\partial I} + \hD_I \hat{X}\\
    \overline{L}_I(\hat{X}) = \frac{\partial X}{\partial I} - \hat{X} \hD_I
\end{gather}
We can then write the second line term-by-term:
\begin{align}
    \hY_W^\dagger \pb{\hat{\tilde{H}}_W}{\hY_1} &= \frac{i}{4}\nabla_R\left(\hLambda_W\right) L_P\left(\hat{C}\right) - \frac{i}{4}\nabla_P\left(\hLambda_W\right)L_R\left(\hat{C}\right)\\
    \hY_1^\dagger \pb{\hat{\tilde{H}}_W}{\hY} &= \frac{i}{4} \hat{C}\left( \nabla_R\left(\hLambda_W\right)\hD_P - \nabla_P\left(\hLambda_W \right)\hD_R \right)\\
    \pb{\hY_1^\dagger}{\hY_W\hLambda_W} &= \frac{i}{4}\overline{L}_R\left(\hat{C}\right) L_P\left(\hLambda_W\right) - \frac{i}{4}\overline{L}_P\left(\hat{C} \right)L_R\left(\hLambda_W\right)\\
    \pb{\hY^\dagger}{\hat{\tilde{H}}_W\hY_1} &= -\frac{i}{4} \hD_RL_P\left( \hLambda_W \hat{C} \right)+ \frac{i}{4}\hD_PL_R\left( \hLambda_W\hat{C}\right)
\end{align}
Finally, for the third line, it will be convenient to define the 2nd order derivatives of the wavefunctions and the covariant energy hessians:
\begin{gather}
    Q_{RR} \equiv \hY_W^\dagger \hY_{RR} = \hD_{RR} + \hD_R^2\\
    Q_{PP} \equiv\hY_W^\dagger \hY_{PP} = \hD_{PP} + \hD_P^2\\\
    Q_{RP} \equiv\hY_W^\dagger \hY_{RP} = \hD_{PR} + \hD_{R}\hD_{P}\\
    L_I^2\left( \hLambda_W\right) \equiv \hLambda_{II} + \hD_{II}\hLambda_W + 2\hD_I\hLambda_I + \hD^2_I\hLambda_W\\
    L_RL_P\left(\hLambda_W\right) \equiv  \hLambda_{RP} + \hD_{PR}\hLambda_W + \hD_P\hLambda_R + \hD_R\hLambda_P + \hD_R\hD_P\hLambda \\
    \nabla_I^2(\hLambda_W) \equiv \hLambda_{II} + \com{\hD_{II}}{\hLambda_W} + 2\com{\hD_I}{\hLambda_I} + \com{\hD_I}{\com{\hD_I}{\hLambda_W}}\\
    \nabla_R \nabla_P \left(\hLambda_W\right) \equiv \hLambda_{RP}  + \com{\hD_{PR}}{\hLambda_W} + \com{\hD_P}{\hLambda_R} + \com{\hD_R}{\hLambda_P} + \com{\hD_R}{\com{\hD_P}{\hLambda_W}}
\end{gather}
We then can succinctly define the terms in the third line of Eq. \ref{eq:lambda-2nd}:
\begin{equation}
    \begin{aligned}
        \hY_W^\dagger \dpb{\hat{\tilde{H}}}{\hY_W} &= \nabla^2_{R}\left(\hLambda_W\right) Q_{PP} + \nabla^2_{P}\left(\hLambda_W\right)Q_{RR} - 2\nabla_R\nabla_P\left(\hLambda_W\right)Q_{RP}\\
        \pb{\hY_W^\dagger}{\pb{\hat{\tilde{H}}}{\hY_W}} &=  -\hD_R L_P\left( \nabla_R\left(\hLambda_W\right) \hD_P - \nabla_P\left(\hLambda_W\right) \hD_R\right) + \hD_PL_R(\nabla_R\left(\hLambda_W\right) \hD_P - \nabla_P\left(\hLambda_W\right) \hD_R) \\
        \dpb{\hY_W^\dagger}{\hY_W \hLambda_W} &= Q^\dagger_{RR}L_P^2\left(\hLambda_W\right) + Q^\dagger_{PP}L^2_R\left(\hLambda_W\right) -2Q^\dagger_{RP}L_R L_P \left( \hLambda_W \right)
    \end{aligned}
\end{equation}
A tractable representation for the total 2nd order correction is therefore:
\begin{equation}
    \begin{aligned}
        \hLambda_2 &= \acom{\hat{K}}{\hLambda_W} - \frac{1}{16} \hat{C}\hLambda_W \hat{C}\\
        &+\frac{i}{2}\left[ \frac{i}{4}\nabla_R\left(\hLambda_W\right) L_P\left(\hat{C}\right) - \frac{i}{4}\nabla_P\left(\hLambda_W\right)L_R\left(\hat{C}\right) \right.  + \frac{i}{4} \hat{C}\left( \nabla_R\left(\hLambda_W\right)\hD_P - \nabla_P\left(\hLambda_W \right)\hD_R \right)\\
        &\qquad + \frac{i}{4}\overline{L}_R\left(\hat{C}\right) L_P\left(\hLambda_W\right) - \frac{i}{4}\overline{L}_P\left(\hat{C} \right)L_R\left(\hLambda_W\right) \left. - \frac{i}{4} \hD_RL_P\left( \hLambda_W \hat{C} \right)+ \frac{i}{4}\hD_PL_R\left( \hLambda_W\hat{C}\right) \right]\\
        &-\frac{1}{8}\left[\nabla^2_{R}\left(\hLambda_W\right) Q_{PP} + \nabla^2_{P}\left(\hLambda_W\right)Q_{RR} - 2\nabla_R\nabla_P\left(\hLambda_W\right)Q_{RP}\right.\\ 
        &\qquad +2\left( -\hD_R L_P\left( \nabla_R\left(\hLambda_W\right) \hD_P - \nabla_P\left(\hLambda_W\right) \hD_R\right) + \hD_PL_R(\nabla_R\left(\hLambda_W\right) \hD_P - \nabla_P\left(\hLambda_W\right) \hD_R) \right)\\
        &\qquad\left. +  Q^\dagger_{RR}L_P^2\left(\hLambda_W\right) + Q^\dagger_{PP}L^2_R\left(\hLambda_W\right) -2Q^\dagger_{RP}L_R L_P \left( \hLambda_W \right)\right]
    \end{aligned}
\end{equation}
\bibliography{mybib,finalbib}

@string{JCP = "{\em J. Chem. Phys.}"}

@string{JPCA = "{\em J. Phys. Chem. A}"}

@string{RMP = "{\em Rev. of Mod. Phys.}"}

@string{PRB= "{\em Phys. Rev. B}"}

@string{CPL = "{\em Chem. Phys. Lett.}"}

@string{ACR = "{\em Acc. Chem. Res.}"}

@string{JCP = "{ J. Chem. Phys.}"}

@string{JPCA = "{ J. Phys. Chem. A}"}

@string{RMP = "{ Rev. of Mod. Phys.}"}

@string{PRB= "{ Phys. Rev. B}"}

@string{CPL = "{ Chem. Phys. Lett.}"}

@string{ACR = "{ Acc. Chem. Res.}"}

@string{JCP = "Journal of Chemical Physics"}

@string{JPCA = "Journal of Physical Chemistry A"}

@string{RMP = "Reviews of Modern Physics"}

@string{PRB = "Physical Review B"}

@string{CPL = "Chemical Physics Letters"}

@string{ACR = "Accounts of Chemical Research"}

@article{PP,
   author = {A. C. Hurley and J. Lennard-Jones and J. A. Pople},
   title = {},
   journal = {Proc. Roy. Soc. A},
   volume = {220},
   pages = {446},
   year = {1953}
}

@book{kubo1991statistical2,
  title={Statistical Physics II: Nonequilibrium Statistical Mechanics},
  author={Kubo, Ryogo and Toda, Morikazu and Hashitsume, Natsuki},
  volume={31},
  series={Springer Series in Solid-State Sciences},
  edition={2nd},
  year={1991},
  publisher={Springer-Verlag},
  address={Berlin, Heidelberg},
  isbn={978-3-540-53833-3}
}

@book{
   martinbook,
   author = {R. M. Martin},
   title = {Electronic Structure},
   publisher = {Cambridge University Press},
   address = {U.K.},
   year = {2004}
}

@article{
   cave:1996:gmh,
   author = {R. J. Cave and M. D. Newton},
   title = {Generalization of the Mulliken-Hush treatment for the calculation of electron transfer matrix elements},
   journal=CPL,
   volume = {249},
   number = {},
   pages = {15-19},
   year = {1996}
}

@article{
   cave:1997:gmh,
   author = {R. J. Cave and M. D. Newton},
   title = {Calculation of electronic coupling matrix elements for ground and excited state electron transfer reactions: Comparison of the generalized Mulliken-Hush and block diagonalization methods},
   journal=JCP,
   volume = {106},
   number = {},
   pages = {9213-9216},
   year = {1997}
}

@article{
   subotnik:2008:boysgmh,
   author = {J. E. Subotnik and S. Yeganeh and R. J. Cave and M. A. Ratner},
   title = {Constructing diabatic states from adiabatic states: Extending generalized Mulliken-Hush to multiple charge centers with Boys localization},
   journal=JCP,
   volume = {129},
   number = {},
   pages = {244101},
   year = {2008},
   note = {}
}

@incollection{
   cederbaum:review:conicalbook,
   author = {L. S. Cederbaum},
   title = {Born-Oppenheimer Approximation and Beyond},
   editor = {W. Domcke and D. R. Yarkony and H. Koppel},
   booktitle = {Conical Intersections: Electronic Structure, Dynamics and Spectroscopy},
   pages = {3-40},
   year = {2004},
   publisher = {World Scientific Publishing Co.},
   address = {New Jersey}
}

@article{
   mbaer:1975:cpl,
   author = {M. Baer},
   title = {Adiabatic and diabatic representations for atom-molecule collisions: Treatment of the collinear arrangement  },
   journal=CPL,
   volume = {35},
   number = {},
   pages = {112},
   year = {1975}
}

@article{
   hsu:2006:tt,
   author = {Z. Q. You and C. P. Hsu and G. Fleming},
   title = {Triplet-Triplet Energy-Transfer Coupling: Theory and Calculation},
   journal=JCP,
   volume = {124},
   number = {},
   pages = {044506},
   year = {2006}
}

@article{
   voityuk:2002:fcd,
   author = {A. A. Voityuk and N. Rosch},
   title = {Fragment charge difference method for estimating donor¿acceptor electronic coupling: Application to DNA $\pi$-stacks},
   journal=JCP,
   volume = {117},
   number = {},
   pages = {5607-5616},
   year = {2002}
}

@article{
   tully:fssh,
   author = {J. C. Tully},
   title = {Molecular dynamics with electronic transitions},
   journal=JCP,
   volume = {93},
   number = {},
   pages = {1061-1071},
   year = {1990}
}

@article{
   kapral:1999:jcp,
   author = {R. Kapral and G. Ciccotti},
   title = {Mixed quantum-classical dynamics},
   journal=JCP,
   volume = {110},
   number = {},
   pages = {8919-8929},
   year = {1999}
}

@article{
   martens:1997:partwig,
   author = {C. C. Martens and J. Y. Fang},
   title = {Semiclassical-limit molecular dynamics on multiple electronic surfaces},
   journal=JCP,
   volume = {106},
   number = {},
   pages = {4918-4930},
   year = {1997}
}

@article{
   coker:2012:iterative,
   author = {P. Huo and D. Coker},
   title = { Consistent schemes for non-adiabatic dynamics derived from partial linearized density matrix propagation },
   journal=JCP,
   volume = {137},
   number = {},
   pages = {22A535},
   year = {2012}
}

@article{
   coker:2008:iterative,
   author = {E. R. Dunkel and S. Bonella and D. Coker},
   title = {Iterative linearized approach to nonadiabatic dynamics},
   journal=JCP,
   volume = {129},
   number = {},
   pages = {114106},
   year = {2008}
}

@article{
   kapral:2010:jcp_pbme,
   author = { A. Nassimi and S. Bonella and R. Kapral},
   title = {Analysis of the quantum-classical liouville equation in the mapping basis},
   journal=JCP,
   volume = {133},
   number = {},
   pages = {134115},
   year = {2010}
}

@article{
   ymrhee:2014:jcp_pbme_new,
   author = {H. W. Kim and  Y. M. Rhee},
   title = { Improving long time behavior of Poisson bracket mapping equation: A non-Hamiltonian approach},
   journal=JCP,
   volume = {140},
   number = {},
   pages = {184106},
   year = {2014}
}

@book{
   tannor:quantumbook,
   author = {D. Tannor},
   title = {Introduction to Quantum Mechanics: A Time-Dependent Perspective},
   publisher = {University Science Books},
   address = {},
   year = {2006}
}

@article{
subotnik:2015:acr,
author = {Subotnik, Joseph E. and Alguire, Ethan C. and Ou, Qi and Landry, Brian R. and Fatehi, Shervin},
title = {The Requisite Electronic Structure Theory To Describe Photoexcited Nonadiabatic Dynamics: Nonadiabatic Derivative Couplings and Diabatic Electronic Couplings},
journal = ACR,
volume = {48},
number = {5},
pages = {1340-1350},
year = {2015},
}

@article{
  yarkony:1996:rmp,
  title = {Diabolical conical intersections},
  author = {Yarkony, David R.},
  journal = RMP,
  volume = {68},
  issue = {4},
  pages = {985--1013},
  numpages = {0},
  year = {1996},
  month = {Oct},
  publisher = {American Physical Society},
  doi = {10.1103/RevModPhys.68.985},
  url = {https://link.aps.org/doi/10.1103/RevModPhys.68.985}
}

@article{
  mead:1992:rmp,
  title = {The geometric phase in molecular systems},
  author = {Mead, C. Alden},
  journal = RMP,
  volume = {64},
  issue = {1},
  pages = {51--85},
  numpages = {0},
  year = {1992},
  month = {Jan},
  publisher = {American Physical Society},
  doi = {10.1103/RevModPhys.64.51},
  url = {https://link.aps.org/doi/10.1103/RevModPhys.64.51}
}

@article{
littlejohn:2022:jcp:parallel,
  title={The parallel-transported (quasi)-diabatic basis},
  author={Littlejohn, Robert and Rawlinson, Jonathan and Subotnik, Joseph},
  journal={The Journal of Chemical Physics},
  volume={157},
  pages={184303},
  year={2022},
  publisher={AIP Publishing}
}

@article{
  nafie:1983:jcp:el_momentum,
  title={Adiabatic molecular properties beyond the Born--Oppenheimer approximation. Complete adiabatic wave functions and vibrationally induced electronic current density},
  author={Nafie, Laurence A},
  journal=JCP,
  volume={79},
  number={10},
  pages={4950--4957},
  year={1983},
  publisher={American Institute of Physics}
}

@book{
   allentildesleybook,
   author = {M. P. Allen and D. J. Tildesley},
   title = {Computer Simulation of Liquids},
   publisher = {Oxford Science Publications},
   address = {New York},
   year = {1987}
}

@article{
   xuezhi:2023:total_ang_bomd,
   author = {Xuezhi Bian and Zhen Tao and Yanze Wu and Jonathan Rawlinson and Robert G. Littlejohn and Joseph E. Subotnik},
   title = {Total Angular Momentum Conservation in Ab Initio Born-Oppenheimer Molecular Dynamics},
   journal=PRB,
   volume = {108},
   number = {},
   pages = {L220304},
   year = {2023},  
   doi= {},
   note={https://dx.doi.org/10.1103/PhysRevB.108.L220304}
}

@article{
coraline:2024:jcp:pssh_conserve,
    author = {Tao, Zhen and Qiu, Tian and Bhati, Mansi and Bian, Xuezhi and Duston, Titouan and Rawlinson, Jonathan and Littlejohn, Robert G. and Subotnik, Joseph E.},
    title = {Practical phase-space electronic Hamiltonians for ab initio dynamics},
    journal = JCP,
    volume = {160},
    number = {12},
    pages = {124101},
    year = {2024},
    month = {03},
    issn = {0021-9606},
    doi = {10.1063/5.0192084},
    url = {https://doi.org/10.1063/5.0192084},
    eprint = {https://pubs.aip.org/aip/jcp/article-pdf/doi/10.1063/5.0192084/19846377/124101\_1\_5.0192084.pdf},
}

@article{
  nafie:1997:jpca_current_density1,
  title={Electron transition current density in molecules. 1. Non-Born- Oppenheimer theory of vibronic and vibrational transitions},
  author={Nafie, Laurence A},
  journal=JPCA,
  volume={101},
  number={42},
  pages={7826--7833},
  year={1997},
  publisher={ACS Publications}
}

@article{
    nafie:1992:vcd,
    author = {Nafie, Laurence A.},
    title = "{Velocity‐gauge formalism in the theory of vibrational circular dichroism and infrared absorption}",
    journal = JCP,
    volume = {96},
    number = {8},
    pages = {5687-5702},
    year = {1992},
    month = {04},
    issn = {0021-9606},
    doi = {10.1063/1.462668},
    url = {https://doi.org/10.1063/1.462668},
    eprint = {https://pubs.aip.org/aip/jcp/article-pdf/96/8/5687/18998695/5687\_1\_online.pdf},
}

@article{
  gross:2015:jcp:vcd_exact_factorization,
  title={Nuclear velocity perturbation theory for vibrational circular dichroism: An approach based on the exact factorization of the electron-nuclear wave function},
  author={Scherrer, Arne and Agostini, Federica and Sebastiani, Daniel and Gross, EKU and Vuilleumier, Rodolphe},
  journal=JCP,
  volume={143},
  number={7},
  year={2015},
  publisher={AIP Publishing}
}

@article{
  case:2008:wigner_review,
  title={Wigner functions and Weyl transforms for pedestrians},
  author={Case, William B},
  journal={American Journal of Physics},
  volume={76},
  number={10},
  pages={937--946},
  year={2008},
  publisher={AIP Publishing}
}

@article{
    bian:2024:pssh_translations_rotations,
    author = {Bian, Xuezhi and Wu, Yanze and Qiu, Tian and Tao, Zhen and Subotnik, Joseph E.},
    title = {A semiclassical non-adiabatic phase-space approach to molecular translations and rotations: Surface hopping with electronic inertial effects},
    journal = {The Journal of Chemical Physics},
    volume = {161},
    number = {23},
    pages = {234114},
    year = {2024},
    month = {12},
    issn = {0021-9606},
    doi = {10.1063/5.0242673},
    url = {https://doi.org/10.1063/5.0242673},
    eprint = {https://pubs.aip.org/aip/jcp/article-pdf/doi/10.1063/5.0242673/20312075/234114\_1\_5.0242673.pdf},
}

@article{
xuezhi:cpr:review:2026,
    author = {Bian, Xuezhi and Duston, Titouan and Bradbury, Nadine and Tao, Zhen and Bhati, Mansi and Qiu, Tian and Wu, Xinchun and Wu, Yanze and Subotnik, Joseph E.},
    title = {The phase-space way to electronic structure theory and subsequently chemical dynamics},
    journal = {Chemical Physics Reviews},
    volume = {7},
    number = {1},
    pages = {011303},
    year = {2026},
    month = {01},
    issn = {2688-4070},
    doi = {10.1063/5.0286240},
    url = {https://doi.org/10.1063/5.0286240},
    eprint = {https://pubs.aip.org/aip/cpr/article-pdf/doi/10.1063/5.0286240/20867104/011303\_1\_5.0286240.pdf}
}

@article{
polkovnikov:2026:pnas,
author = {Bernardo Barrera  and Daniel P. Arovas  and Anushya Chandran  and Anatoli Polkovnikov },
title = {The moving Born–Oppenheimer approximation},
journal = {Proceedings of the National Academy of Sciences},
volume = {123},
number = {7},
pages = {e2507816123},
year = {2026},
doi = {10.1073/pnas.2507816123},
URL = {https://www.pnas.org/doi/abs/10.1073/pnas.2507816123},
eprint = {https://www.pnas.org/doi/pdf/10.1073/pnas.2507816123}}

@book{
   born:huang,
   author = {Max Born and Kun Huang},
   title = {Dynamical Theory of Crystal Lattices},
   publisher = {Oxford University Press},
   address = {Oxford},
   year = {1954}
}

@misc{wiggins_2026_solvable_models_reveal_born_oppenheimer,
      title={What solvable models reveal about Born-Oppenheimer, Born-Huang, and exact factorization}, 
      author={Stephen Wiggins},
      year={2026},
      eprint={2608.08668},
      archivePrefix={arXiv},
      primaryClass={quant-ph},
      url={https://arxiv.org/abs/2608.08668}, 
}

@article{Zaidi-electron_transfer-2026,
    author = {Zaidi, Zain and Bian, Xuezhi and Subotnik, Joseph E.},
    title = {Electron transfer, diabatic couplings, and vibronic energy gaps in a phase space electronic structure framework},
    journal = {The Journal of Chemical Physics},
    volume = {164},
    number = {19},
    pages = {194112},
    year = {2026},
    month = {05},
    issn = {0021-9606},
    doi = {10.1063/5.0325462},
    url = {https://doi.org/10.1063/5.0325462},
}

@article{tao_basis-free_2025,
	title = {A {Basis}-{Free} {Phase} {Space} {Electronic} {Hamiltonian} {That} {Recovers} {Beyond} {Born}-{Oppenheimer} {Electronic} {Momentum} and {Current} {Density}},
    journal = {Jour. Chem. Phys.},
    year = {2025},
	author = {Tao, Zhen and Qiu, Tian and Bian, Xuezhi and Duston, Titouan and Bradbury, Nadine and Subotnik, Joseph E},
}

@article{bian_PS-vibration_2025,
author = {Bian, Xuezhi and Khan, Cameron and Duston, Titouan and Rawlinson, Jonathan and Littlejohn, Robert G. and Subotnik, Joseph E.},
title = {A Phase-Space View of Vibrational Energies without the Born–Oppenheimer Framework},
journal = {Journal of Chemical Theory and Computation},
volume = {21},
number = {6},
pages = {2880-2893},
year = {2025},
doi = {10.1021/acs.jctc.4c01294},
URL = {https://doi.org/10.1021/acs.jctc.4c01294}
}

@article{wu_exact-vibration_2025,
author = {Wu, Xinchun and Bian, Xuezhi and Rawlinson, Jonathan and Littlejohn, Robert G. and Subotnik, Joseph E.},
title = {Recovering Exact Vibrational Energies within a Phase Space Electronic Structure Framework},
journal = {Journal of Chemical Theory and Computation},
volume = {21},
number = {19},
pages = {9470-9482},
year = {2025},
doi = {10.1021/acs.jctc.5c00956},
URL = {https://doi.org/10.1021/acs.jctc.5c00956}}

@article{tao-ROA-2026,
    author = {Tao, Zhen and Bhati, Mansi and Subotnik, Joseph E.},
    title = {Non-resonant Raman optical activity from phase-space electronic structure theory},
    journal = {APL Computational Physics},
    volume = {2},
    number = {2},
    pages = {026101},
    year = {2026},
    month = {04},
    issn = {3066-0017},
    doi = {10.1063/5.0315696},
    url = {https://doi.org/10.1063/5.0315696}
    }

@article{tao-VCD-2024,
    author = {Tao, Zhen and Duston, Titouan and Pei, Zheng and Shao, Yihan and Rawlinson, Jonathan and Littlejohn, Robert and Subotnik, Joseph E.},
    title = {An electronic phase-space Hamiltonian approach for electronic current density and vibrational circular dichroism},
    journal = {The Journal of Chemical Physics},
    volume = {161},
    number = {20},
    pages = {204107},
    year = {2024},
    month = {11},
    issn = {0021-9606},
    doi = {10.1063/5.0233618},
    url = {https://doi.org/10.1063/5.0233618}
}

@article{duston-VCD-2024,
    author = {Duston, Titouan and Tao, Zhen and Bian, Xuezhi and Bhati, Mansi and Rawlinson, Jonathan and Littlejohn, Robert G. and Pei, Zheng and Shao, Yihan and Subotnik, Joseph E.},
    title = {A Phase-Space Electronic Hamiltonian For Vibrational
Circular Dichroism},
    journal = {Journal of Chemical Theory and Computation},
    volume = {20},
    number = {18},
    pages = {7904-7921},
    year = {2024},
    month = {09},
    issn = {1549-9618},
    doi = {10.1021/acs.jctc.4c00662},
    url = {https://doi.org/10.1021/acs.jctc.4c00662}
}

@article{born-huang_dynamical_1955,
    author = {Born, Max and Huang, Kun and Lax, M.},
    title = {Dynamical Theory of Crystal Lattices},
    journal = {American Journal of Physics},
    volume = {23},
    number = {7},
    pages = {474-474},
    year = {1955},
    month = {10},
    issn = {0002-9505},
    doi = {10.1119/1.1934059},
    url = {https://doi.org/10.1119/1.1934059}
}

@article{littlejohn-flynn_perturbation-theory_1991,
  title = {Geometric phases in the asymptotic theory of coupled wave equations},
  author = {Littlejohn, Robert G. and Flynn, William G.},
  journal = {Phys. Rev. A},
  volume = {44},
  issue = {8},
  pages = {5239--5256},
  numpages = {0},
  year = {1991},
  month = {Oct},
  publisher = {American Physical Society},
  doi = {10.1103/PhysRevA.44.5239},
  url = {https://link.aps.org/doi/10.1103/PhysRevA.44.5239}
}

@article{heller-wignerPS-1976,
    author = {Heller, Eric J.},
    title = {Wigner phase space method: Analysis for semiclassical applications},
    journal = {The Journal of Chemical Physics},
    volume = {65},
    number = {4},
    pages = {1289-1298},
    year = {1976},
    month = {08},
    issn = {0021-9606},
    doi = {10.1063/1.433238},
    url = {https://doi.org/10.1063/1.433238}
}

@article{borgis-model-2006,
title = {Generating approximate Wigner distributions using Gaussian phase packets propagation in imaginary time},
journal = {Chemical Physics Letters},
volume = {423},
number = {4},
pages = {390-394},
year = {2006},
issn = {0009-2614},
doi = {https://doi.org/10.1016/j.cplett.2006.04.007},
url = {https://www.sciencedirect.com/science/article/pii/S0009261406004787},
author = {Dana Codruta Marinica and Marie-Pierre Gaigeot and Daniel Borgis}
}

@article{MSW-1986,
  title = {Realizations of Magnetic-Monopole Gauge Fields: Diatoms and Spin Precession},
  author = {Moody, John and Shapere, A. and Wilczek, Frank},
  journal = {Phys. Rev. Lett.},
  volume = {56},
  issue = {9},
  pages = {893--896},
  numpages = {0},
  year = {1986},
  month = {Mar},
  publisher = {American Physical Society},
  doi = {10.1103/PhysRevLett.56.893},
  url = {https://link.aps.org/doi/10.1103/PhysRevLett.56.893}
}

@misc{bhati-MSW-2026,
      title={Electronic Structure in a Phase Space, non-Born-Oppenheimer Framework: Geometric Forces and Moody-Shapere-Wilzcek Revisited}, 
      author={Mansi Bhati and D. Vale Cofer-Shabica and Jonathan I. Rawlinson and Robert G. Littlejohn and Joseph Subotnik and Nadine C. Bradbury},
      year={2026},
      eprint={2605.27053},
      archivePrefix={arXiv},
      primaryClass={physics.chem-ph},
      url={https://arxiv.org/abs/2605.27053}, 
}

@article{peng-SR-2026,
    author = {Peng, Linqing and Duston, Titouan and Bradbury, Nadine and Bhati, Mansi and Tao, Xuecheng and Rosen, Michael B. and Subotnik, Joseph E.},
    title = {Conceptual Shift in
Our Understanding of Degenerate
Radical Spin Systems: Spin-Rotation Coupling Turned on Its Head},
    journal = {Journal of the American Chemical Society},
    year = {2026},
    month = {08},
    issn = {0002-7863},
    doi = {10.1021/jacs.6c05615},
    url = {https://doi.org/10.1021/jacs.6c05615}
}

\end{document}